\documentclass{article}

\usepackage{neurips_data_2024}

\usepackage{amsmath} 

\usepackage[utf8]{inputenc} 
\usepackage[T1]{fontenc}    
\usepackage{hyperref}       
\usepackage{url}            
\usepackage{booktabs}       
\usepackage{amsfonts}       
\usepackage{nicefrac}       
\usepackage{microtype}      
\usepackage{xcolor}         
\usepackage{graphicx}
\usepackage{multirow}
\usepackage{subcaption}
\usepackage{caption}
\usepackage{array}
\usepackage{afterpage}

\usepackage{pdfpages}
\usepackage{geometry}
\usepackage{natbib}  
\title{UniEntrezDB: Large-scale Gene Ontology Annotation Dataset and Evaluation Benchmarks with Unified Entrez Gene Identifiers}

\author{%
  Yuwei Miao$^{1}$, Yuzhi Guo$^{1}$, Hehuan Ma$^{1}$, Jingquan Yan$^{1}$, Feng Jiang$^{1}$ \\ \textbf{Weizhi An}$^{1}$, \textbf{Jean Gao}$^{1}$, \textbf{Junzhou Huang}$^{1}$\thanks{Corresponding author: jzhuang@uta.edu}  \\
  $^{1}$ Department of Computer Science and Engineering, University of Texas at Arlington, Arlington, TX \\
}

\begin{document}

\maketitle

\begin{abstract}
Gene studies are crucial for fields such as protein structure prediction, drug discovery, and cancer genomics, yet they face challenges in fully utilizing the vast and diverse information available. Gene studies require clean, factual datasets to ensure reliable results. Ontology graphs, neatly organized domain terminology graphs, provide ideal sources for domain facts. However, available gene ontology annotations are currently distributed across various databases without unified identifiers for genes and gene products. To address these challenges, we introduce Unified Entrez Gene Identifier Dataset and Benchmarks (UniEntrezDB), the first systematic effort to unify large-scale public Gene Ontology Annotations (GOA) from various databases using unique gene identifiers. UniEntrezDB includes a pre-training dataset and four downstream tasks designed to comprehensively evaluate gene embedding performance from gene, protein, and cell levels, ultimately enhancing the reliability and applicability of LLMs in gene research and other professional settings.

\end{abstract}

\section{Introduction}
Understanding the functions of genes and their products is critical for various research fields, such as protein structure prediction \cite{gligorijevic2021structure, bernhofer2021predictprotein}, protein design \cite{ferruz2022controllable}, drug discovery \cite{fiorentino2023native, lu2023progressive}, cancer genomics and oncology research \cite{stourac2024predictonco}, and AI-driven gene editing technologies \cite{dixit2024advancing, pun2023ai}. Recent advancements in LLMs show promise for incorporating large-scale domain knowledge documented as text descriptions into real-world applications \cite{chen2023genept,qu2024rise}. However, several challenges hinder the combination of text and professional domain applications.

Ontology, as an organized directed acyclic terminology graph, manages and structures comprehensive domain terminologies with clear definitions. To ensure that research and studies are built on a coherent and evolving foundation of widely accepted terms, the scientific community has established various ontology projects. These projects aim to standardize the terminology for describing functionalities and traits across disciplines. Meanwhile, ontology programs continuously update terminologies and add annotations alongside domain advancements, making them ideal sources for formal knowledge graphs to represent explicit factual domain knowledge \cite{gene2004gene, gene2019gene,ashburner2000gene, shang2024ontofact}. In gene study, Gene Ontology (GO) focuses on describing the functions of genes and gene products, such as RNA and protein \cite{gene2004gene}. Cell Ontology (CL) categorizes and details cell types across various organisms, fostering a unified understanding of cell functions and characteristics \cite{bard2005ontology}. Uber-anatomy Ontology (Uberon) provides a cross-species perspective on anatomy, categorizing entities by structure, function, and lineage \cite{mungall2012uberon}. Chemical Entities of Biological Interest (ChEBI) explores chemical substances in terms of their biological roles and relationships \cite{degtyarenko2007chebi}. Protein Ontology (PRO) defines and contextualizes protein-related entities, enhancing our understanding of their interrelations \cite{natale2010protein}. Phenotype and Trait Ontology (PATO) focuses on the qualities that define phenotypes, offering insights into the attributes that characterize organisms \cite{kohler2019expansion}. This study aims to demonstrate the utility and efficacy of ontology applications across various fields by utilizing the Gene Ontology (GO) database as a prime example. By exploring how structured biological knowledge within GO can be effectively applied, we intend to provide a template that can be adapted to other ontological frameworks.

To explore and evaluate the performance of ontology in bridging the gap between professional text descriptions and real-world applications, we introduce Unified Entrez Gene Identifier Dataset and Benchmarks (UniEntrezDB). To the best of our knowledge, UniEntrezDB is the first systematic attempt to unify large-scale public Gene 
Ontology Annotations (GOA) from various databases and downstream benchmarks with unique gene identifiers. UniEntrezDB includes a pre-train dataset of GOA and four downstream tasks among gene, protein, and cell levels. Our results show that GOA enriches gene information by providing insights into functional aspects and improves the performance of various biological downstream tasks at different levels.

\section{Background}

We start by providing an introduction to the biological relationships between genes and their products, offering a foundational understanding of how these elements interact. Next, we delve into the structure of the GO, explaining its organization and the methods used for annotating genes and their products using the GO framework. This background information collectively details the mechanisms by which genes and their products are linked, organized, and annotated, offering essential insights into gene function across biological and computational research.

\subsection{Gene and Gene Products}
In biology, genes are specific sequences of nucleotides within DNA that control hereditary characteristics, primarily through regulating the synthesis of the RNA and protein products \cite{johnson2002molecular}. Following the central dogma, genes within DNA are transcribed into messenger RNA (mRNA), which are then translated into gene products such as proteins. Given a specific gene product, either protein or RNA, we can always trace it back to the gene corresponding to the product’s synthesis process \cite{Lodish2000cellbio}. Given the trackable relationship from gene products back to the original gene, a comprehensive description of a gene’s functionality requires not only the gene itself but all the functions of its products \cite{Gerstein2007df}. 

\begin{figure}[!t]
  \centering
  \includegraphics[width=0.75\linewidth]{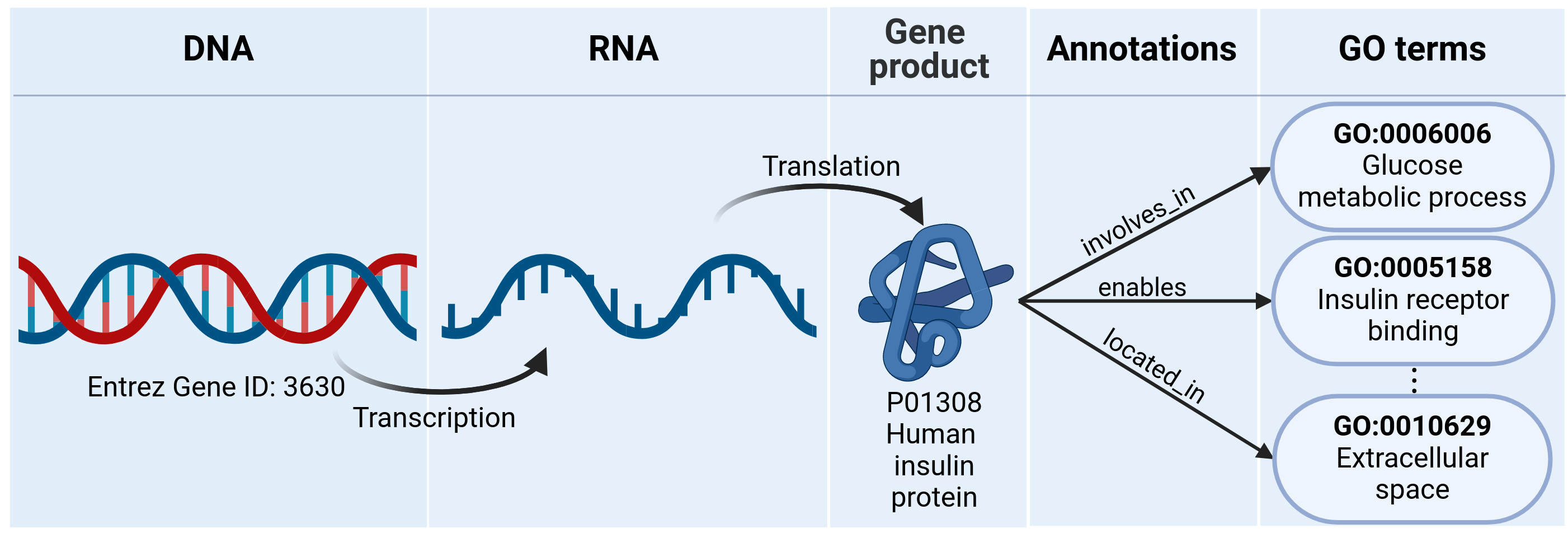}
  \caption{Illustration of the transcription, translation, and GO annotation process. Gene 3630\cite{maglott2005entrez} transcribes into mRNA then translates into Human insulin protein\cite{uniprot2019uniprot} (UniProtKB id: P01308). GO terms are annotated with functions, i.e., Human insulin protein \textit{enables} insulin receptor binding. } 
\label{fig111}
\vspace{-10pt}
\end{figure}

Over the past decades, extensive research has been conducted worldwide to explore the functionalities of genes and their products, and some databases have systematically organized the results with the functionalities annotated using GO. Leveraging the unique gene identifier assigned to each gene in the Entrez Gene system \cite{maglott2005entrez}, we can link a gene identifier to the functionalities of its gene products.  Figure \ref{fig111} illustrates the relationships between gene and gene products and one example of how one gene identifier can link to its product’s GOA.

\subsection{Directed Acyclic Gene Ontology Graph}
\begin{figure}[!ht]
    \centering
    \begin{subfigure}[b]{0.45\textwidth}
        \centering
        \includegraphics[width=\textwidth]{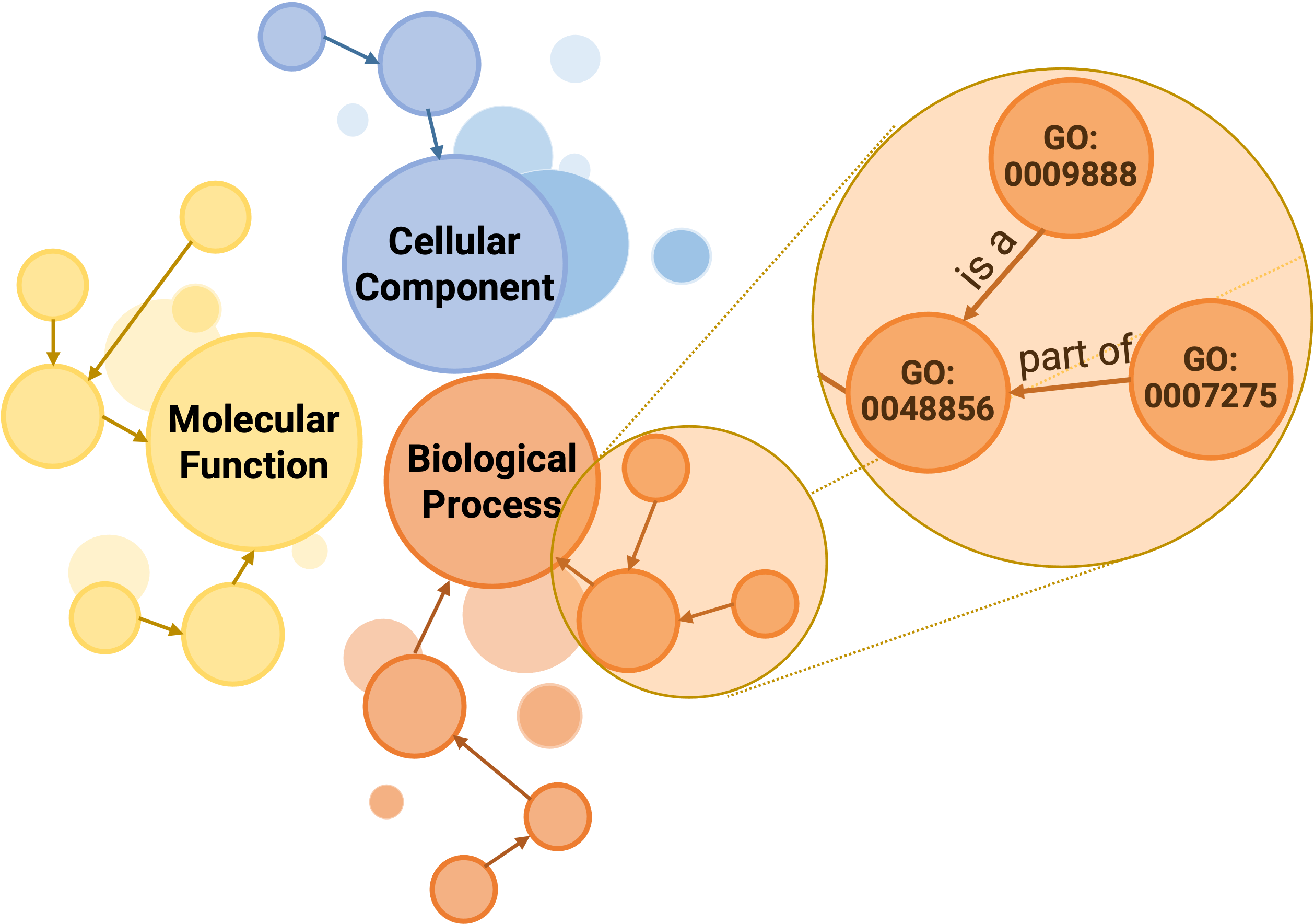}
        \caption{}
        \label{fig:fig2a}
    \end{subfigure}
    \hfill
    \begin{subfigure}[b]{0.45\textwidth}
        \centering
        \includegraphics[width=\textwidth]{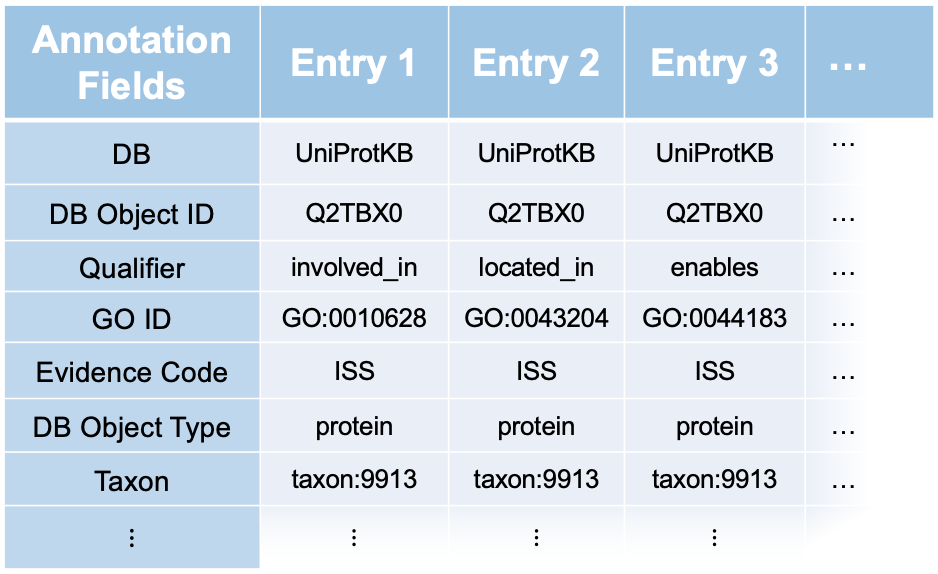}
        \caption{}
        \label{fig:fig2b}
    \end{subfigure}
    
    \caption{\textbf{(a)} Illustration of the GO DAG. Zoom in shows the edge relationship between GO terms: GO:000988 is a GO:0048856 ("embryonic pattern specification" is a "anatomical structure development"), GO:0007275 part of GO:0048856 ("multicellular organism development" part of "anatomical structure development") \textbf{(b)} Examples of GOA in GAF format.}
    \label{fig:combined}
    \vspace{-10pt}
\end{figure}

As shown in Figure \ref{fig:fig2a}, the GO is structured as a Directed Acyclic Graph (DAG) with nodes representing gene functional terminologies and directed edges representing relationships between these terms (is\_a, part\_of, regulates, positively\_regulates, and negatively\_regulates). 

Let \( G = (V, E) \) represent the GO graph, where \( V \) is the set of nodes representing GO terms and \( E \) is the set of directed edges \( (u, v) \) where \( u, v \in V \). The graph \( G \) is a DAG if and only if for all \( v \in V \), there does not exist a path \( P \) such that \( P \) starts and ends at \( v \).
\[
\forall v \in V, \, \neg \exists (v_1, v_2, \ldots, v_k) \text{ such that } v_1 = v_k = v \text{ and } (v_i, v_{i+1}) \in E \text{ for all } 1 \leq i < k.
\]
The GO DAG has three primary root nodes corresponding to different types of gene and gene product functional annotations: Biological Process (BP), Cellular Function (CF), and Molecular Function (MF). These root nodes are represented by \( V_{\text{BP}}, V_{\text{CC}}, V_{\text{MF}} \). Each non-obsolete node \( v \in V \) exists at least one path to one of the root nodes:
\[
\forall v \in V, \, \exists \, (v, v_1, \ldots, v_n) \text{ such that } v_n \in \{V_{\text{BP}}, V_{\text{CC}}, V_{\text{MF}}\} \text{ and } (v_i, v_{i+1}) \in E \text{ for all } 0 \leq i < n.
\]

\subsection{GAF and Gene Ontology Annotation}
GOA provides comprehensive and high-quality annotation of gene products using Gene Ontology. GOA ensures that the functional information about genes is consistent and can be integrated across various species and databases. GOA saves in a standardized format: Gene Annotation File (GAF). GAF is a tab-delimited file format (Figure \ref{fig:fig2b}) used for representing gene annotations to the Gene Ontology. Each line in a GAF file represents a single annotation and includes various fields that provide detailed information. These fields include the database from which the annotation is sourced and a unique database-specific identifier for the gene or gene product. The taxon IDs that represent the organism are also provided\cite{schoch2020ncbi, benson2000genbank}. 
Therefore, we utilize the database identifier and the taxon ID from NCBI Taxonomy to systemically map genes and their corresponding products to the Gene EntrezID.

Additionally, GOA includes a qualifier to indicate the relationship between biological objects and the annotated functional GO term. For each annotation, there are references and the evidence code indicates the confidence and type of references supporting this annotation, which provide trackable sources and reliability for each annotation.

\section{Related Work}

We discuss related work from three aspects: 1) traditional use of GO and pathway information, 2) gene embedding and ontology-related models, and 3) existing biological evaluation benchmarks.

\subsection{Gene Enrichment Analysis}
Gene enrichment analysis identifies over-represented sets of genes or proteins in a dataset, revealing associations with specific biological processes or pathways. This analysis helps simplify genomic data interpretation, aiding the discovery of biomarkers and advancing disease understanding. 
Commonly used gene enrichment analysis methods include Gene Set Enrichment Analysis (GSEA), a computational method that identifies significant differences in gene sets between two biological states \cite{subramanian2005gene} DAVID (Database for Annotation, Visualization, and Integrated Discovery) \cite{huang2009systematic} and Enrichr \cite{chen2013enrichr}. DAVID provides functional annotation tools for large gene lists, and Enrichr is an interactive tool for visualizing the collective functions of gene lists.

Although enrichment analysis is a widely used method for identifying biological functions or pathways that are statistically prominent in a specific set of genes,  the large-scale unified ground truth annotation datasets are not fully leveraged. With the development of deep learning, these valuable GOA present an opportunity to derive more meaningful and impactful insights.

\subsection{Gene Embedding Models and Ontology-Based Methods}
Gene Embedding Models generate embeddings for genes or gene products from diverse sources of gene-related data. Gene2Vec \cite{du2019gene2vec}, a data-driven approach, generates distributed representations of gene names based on their co-expression patterns from GEO \cite{barrett2012ncbi}. GeneBERT\cite{mo2021multi}, GeneMask \cite{roy2023genemask}, DNABERT \cite{ji2021dnabert}, DNABERT-2\cite{zhou2023dnabert2}, and LOGO \cite{yang2022integrating} are gene sequence-based methods that generate effective gene embedding. Moreover, the gLM \cite{hwang2024genomic} generates contextualized protein embeddings that capture both genomic context and protein sequences. 

Efforts have been made to integrate ontology into deep learning methods to improve model reliability and performance. To enhance the solidity of LLMs, Ontofact \cite{shang2024ontofact} introduces an ontology-driven reinforcement learning framework to detect and mitigate factual errors in large language models by generating error-prone test cases using knowledge graphs.  OntoProtein\cite{zhang2022ontoprotein}, a framework that combines GO knowledge into protein language models, uses contrastive learning and knowledge-aware negative sampling to enhance protein representation generation. 

\subsection{Benchmarks for Gene and Gene Products}
Besides the gene enrichment analysis, the gene-level tasks include Gene-Gene Interaction (GGI) prediction such as the GGI task provided by Gene2Vec \cite{du2019gene2vec}. There are various benchmark tasks for protein, the most widely analyzed gene product. Protein-Protein Interaction (PPI) task predicts the interaction types between proteins, STRING \cite{szklarczyk2019string} provides the common and stable benchmark for PPI.  TAPE \cite{rao2019evaluating} evaluates protein embeddings across five tasks: Secondary Structure Prediction, Contact Prediction, Remote Homology Detection, Fluorescence Prediction, and Stability Prediction, these tasks indicate the effectiveness of self-supervised pretraining in enhancing the performance of protein modeling. Single-cell type annotation tasks involve analyzing gene expression levels to identify different cell types. Notable benchmarks include hPancreas \cite{baron2016single,lawlor2017single, muraro2016single, segerstolpe2016single, xin2016rna, chen2023transformer}, Immune Human \cite{luecken2022benchmarking}, and Zheng68k \cite{zheng2017massively}. 

Although these benchmarks utilize various input types, such as gene sequences, protein sequences, or RNA-seq gene expression levels, they can be unified via Gene EntrezID. To comprehensively evaluate gene embedding and model performance, we unify various benchmarks with Gene EntrezID and provide four tasks across gene, protein, and cell levels.

\begin{figure}[!t]
  \centering
  \includegraphics[width=0.95\linewidth]{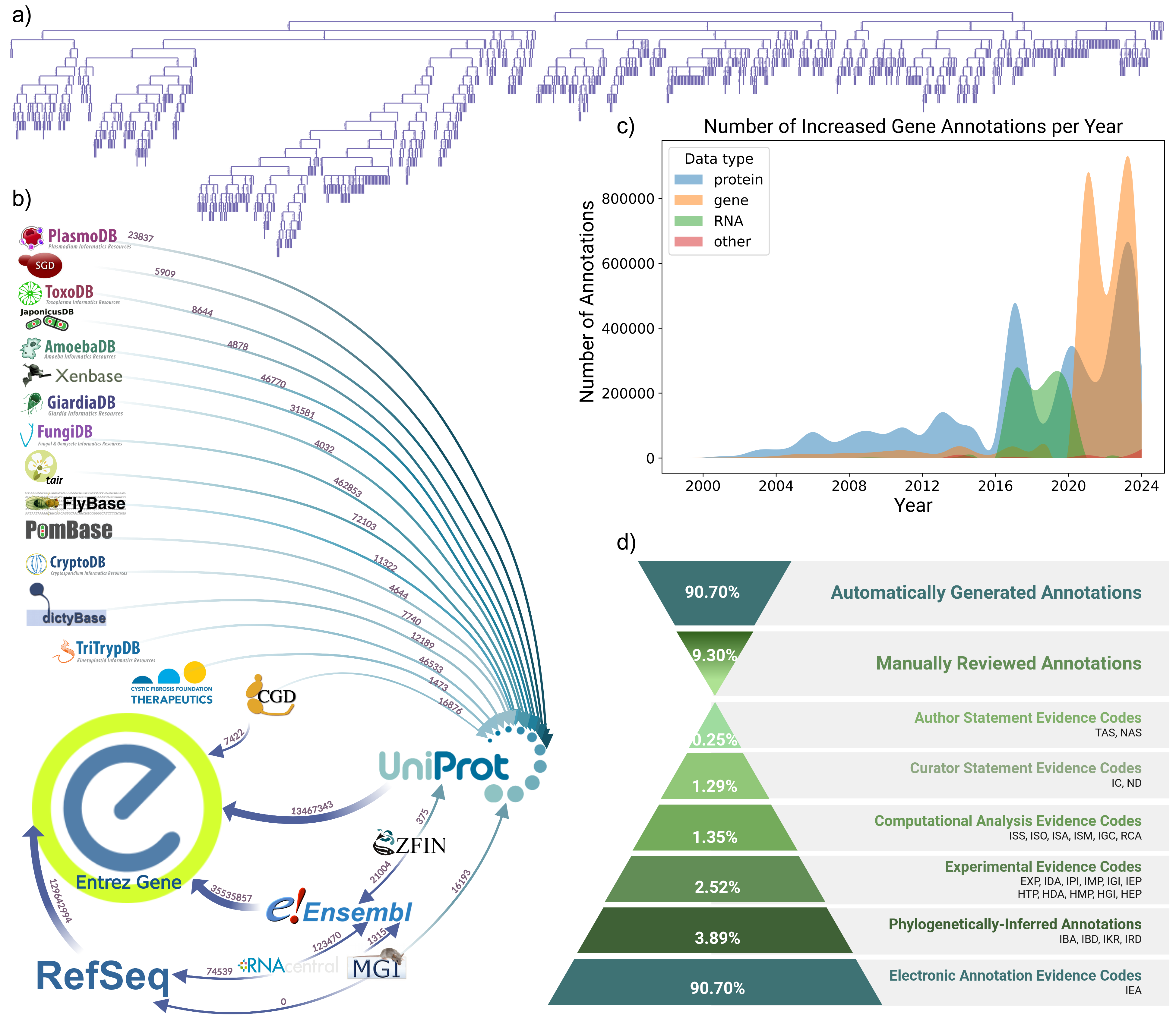}
  \caption{Statistics information of UniEntrezGOA Dataset. More details are available in Appendix. 
  \textbf{a)} Phylogenetic Tree of over 1000 species available in UniEntrezGOA Manually Reviewed Annotations. 
  \textbf{b)} ID mapping procedure between different databases. The numbers on the arrow indicate the number of IDs mapped successfully from the source database to the target database. 
  \textbf{c)} The distribution of increased manually reviewed GOA each year for each gene and gene product.
  \textbf{d)} According to the GO official website, there are six categories of Evidence Codes. Only the Electronic Annotation Evidence Code IEA is not manually reviewed annotations.
  }
  \label{fig3:statistic}
\vspace{-10pt}
\end{figure}

\section{Datasets Construction}
Unified gene identifiers are crucial in gene-related fields, as many studies tend to refer to genes by their names, which can be ambiguous. At first, gene names are determined by their chromosomal locations. However, as research on certain genes deepened, their names were sometimes changed to abbreviations of phrases related to their functions. This evolution in terminology has led to multiple synonyms or historical names for the same gene, complicating the retrieval of gene-related textual information. For instance, Gene2Vec \cite{du2019gene2vec} generates gene co-expression pairs using RNA-seq data and then utilizes skip-gram to train a distributed gene embedding based on gene names. Yet, using gene names as gene identifiers is problematic, resulting in Gene2Vec producing different embedding for the same gene. For example, CXorf22, CHDC2, and CXorf30 all refer to the gene with EntrezID 286464, but the Gene2Vec method produced distinct embedding for this single gene (average pairwise cosine similarity of 42.98\%). Therefore, adopting a unified gene identifier to represent genes and avoid ambiguity is essential.

Our proposed UniEntrezDB comprehensively and systematically unifies public GOA from various databases and gene information from motley benchmarks, incorporating unique gene identifiers for consistency. UniEntrezDB is structured into two primary components: the Gene Ontology Annotation Dataset and the Evaluation Benchmarks. The Gene Ontology Annotation Dataset consolidates annotations from multiple public databases, ensuring a rich and diverse collection of annotations across species. The Evaluation Benchmarks comprise four tasks, designed to assess the performance of gene embedding across different contexts and applications.

\subsection{UniEntrez Gene Ontology Annotation Dataset}
We organize the annotation data across 21 databases and provide a comprehensive dataset: UniEntrezGOA. Dataset statistics are detailed in the Appendix. GO annotations are designed to label the functions among genes and gene products. These functional annotations are associated with specific categories including DNA, RNA, and Protein. Since each gene product corresponds to a unique gene, we have aligned genes and their products to a single gene identifier: Gene EntrezID \cite{maglott2005entrez}. 
We illustrate the ID Unifying process in Figure \ref{fig3:statistic} b) with the logo of each database. As shown, within the aligned 21 databases, RefSeq\cite{o2016reference, pruitt2007ncbi}, Ensembl\cite{martin2023ensembl}, and UniProtKB\cite{uniprot2019uniprot} are the three largest ID systems besides Entrez Gene ID. Most small databases provide the ID mapping files between database-specific IDs and these three largest databases and then convert them to Entrez Gene ID. Over 1,000 species with manually annotated GOA are included in UnientrezGOA, and we plot the taxonomy phylogenetic tree of all these species in Figure \ref{fig3:statistic} a) for demonstration. A full Phenogenetic tree with detailed information is in Appendix. UniEntrezGOA is up to date with current scientific progress, since the GOAs are added more frequently in recent years, as shown in Figure \ref{fig3:statistic} c), which visualizes the increased annotations per year. It then becomes a promising and stable source of data in the future. The distribution of the different Evidence Codes is illustrated in Figure \ref{fig3:statistic} d). The manually reviewed annotation is the reliable ground truth gene and gene product functional annotation. Automatically generated annotations are also good supplementary for the species or gene products that are not well studied or documented.

\subsection{UniEntrez Evaluation Benchmarks}
We deploy four tasks as the UniEntrez Evaluation Benchmarks:
Pathway Co-present Prediction, Functional Gene Interaction Prediction, Protein-Protein Interaction, and Single-Cell Type Annotation using single-cell RNA-seq data. These tasks evaluate the performance of gene embedding models with unified Gene EntrezID across gene, protein, and cell levels. 

\begin{figure}[!ht]
    \centering
    \begin{subfigure}[b]{0.42\textwidth}
        \centering
        \includegraphics[width=\textwidth]{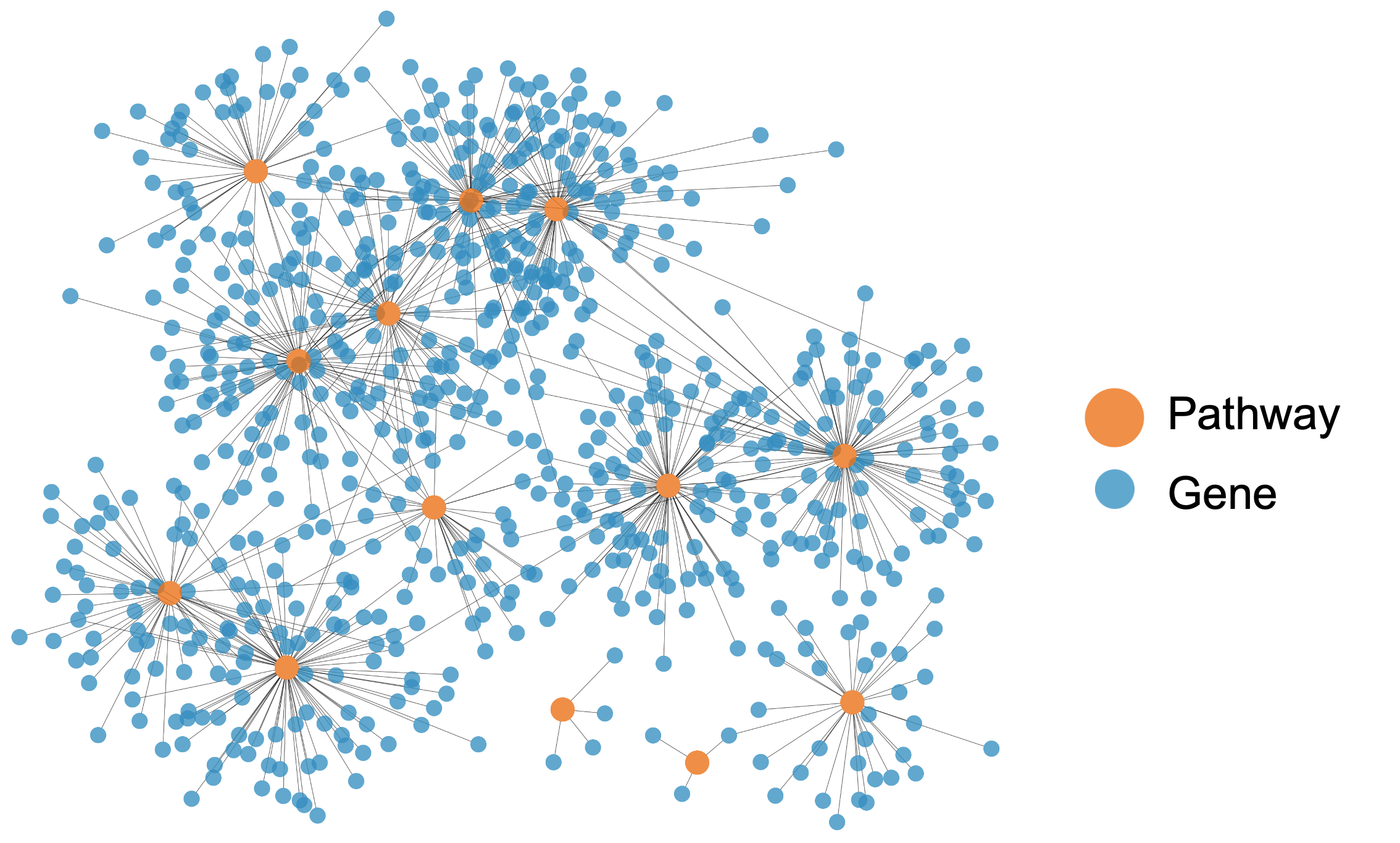}
        \caption{}
        \label{fig4:a}
    \end{subfigure}
    \hfill
    \begin{subfigure}[b]{0.45\textwidth}
        \centering
        \includegraphics[width=\textwidth]{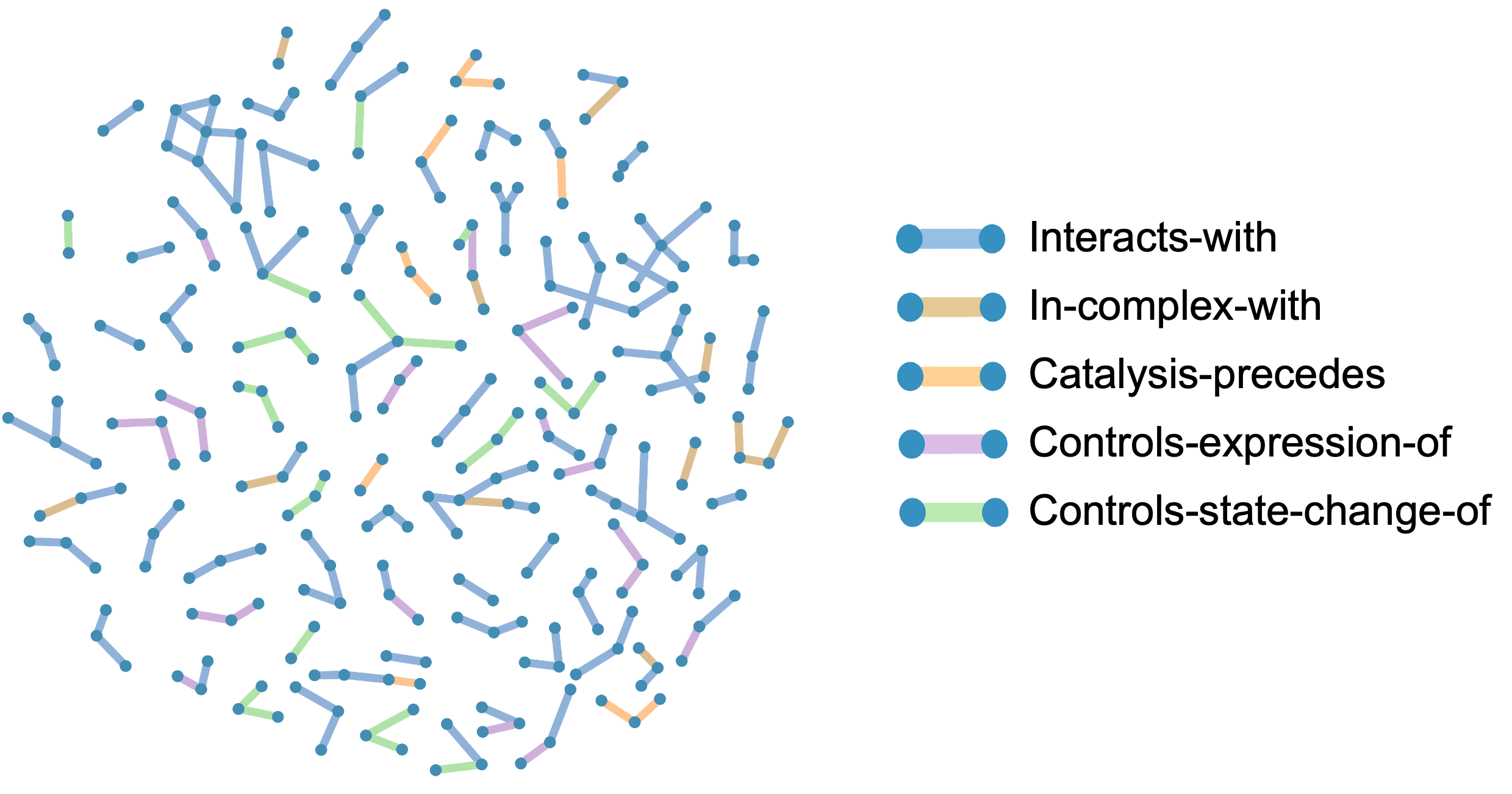}
        \caption{}
        \label{fig4:b}
    \end{subfigure}
    \vspace{1em}
    \begin{subfigure}[b]{0.45\textwidth}
        \centering
        \includegraphics[width=\textwidth]{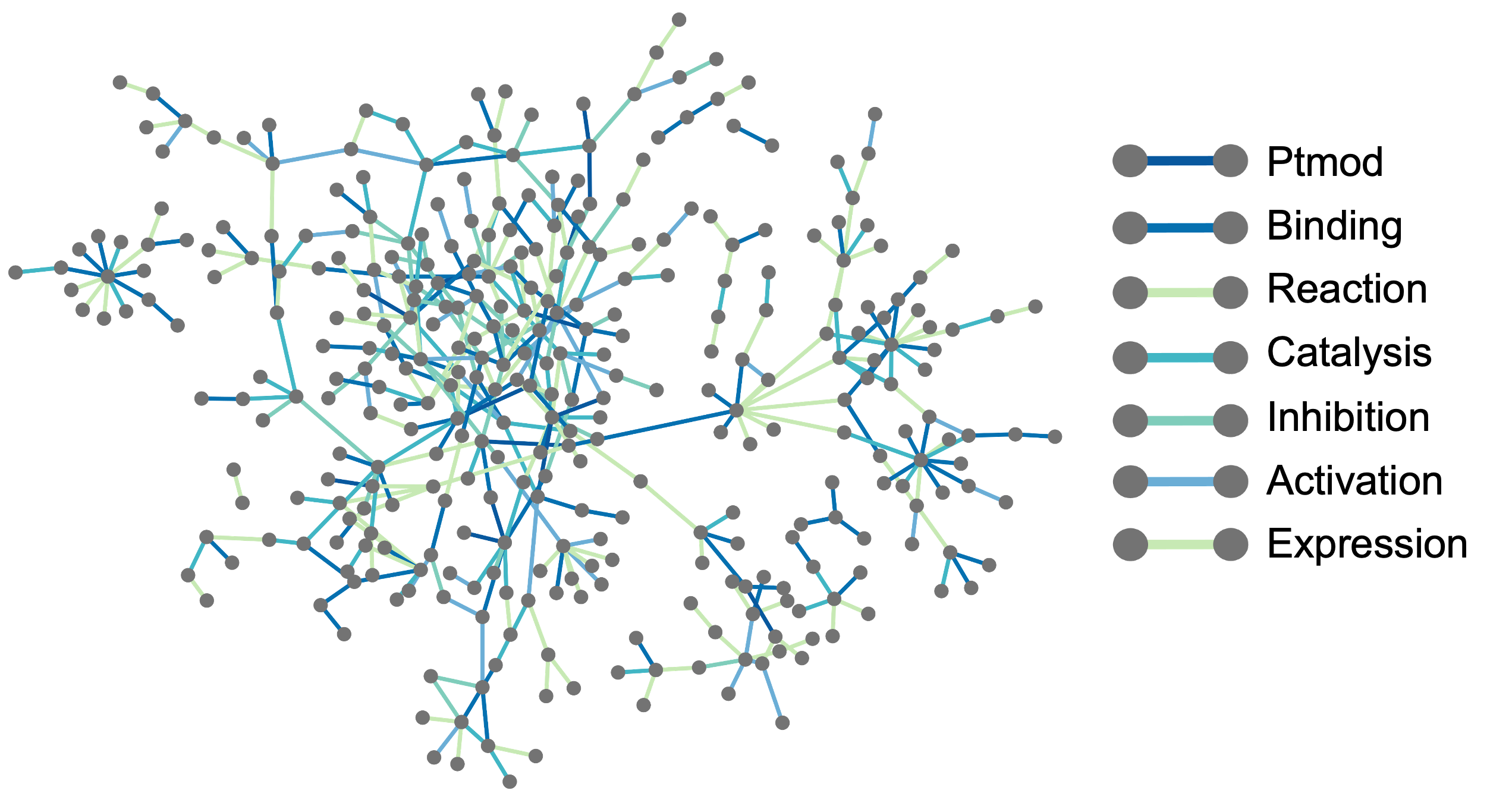}
        \caption{}
        \label{fig4:c}
    \end{subfigure}
    \hfill
    \begin{subfigure}[b]{0.42\textwidth}
        \centering
        \includegraphics[width=\textwidth]{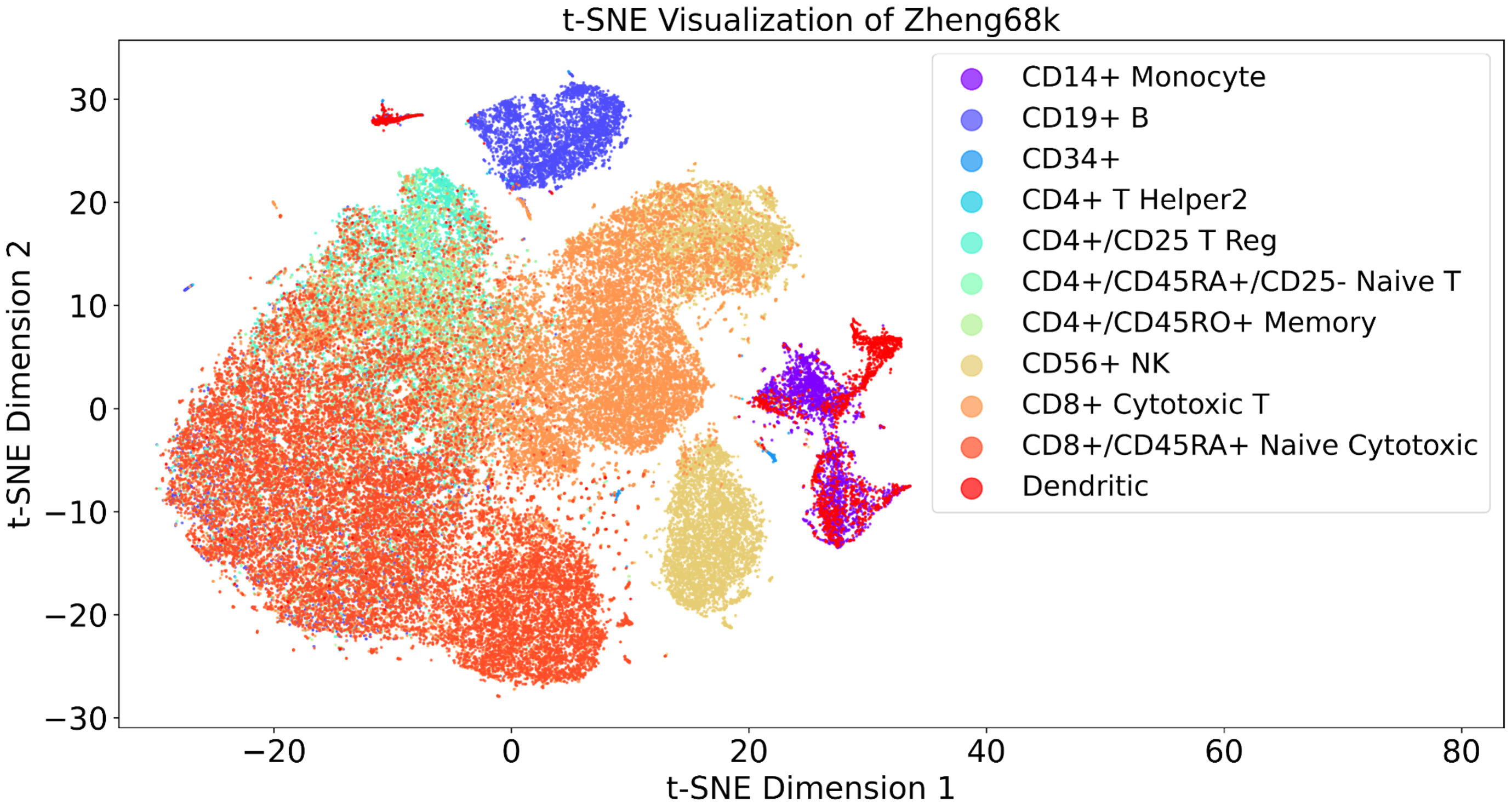}
        \caption{}
        \label{fig4:d}
    \end{subfigure}
    
   \caption{Benchmark Illustration. 
  (a) Pathway Co-present Prediction: The relationship between pathways (Orange) and genes (Blue) in Msigdb\cite{liberzon2011molecular}. Edges form between the pathway and the gene belongs to the pathway. 
  (b) Functional Gene Interaction Prediction: PathwayCommons \cite{wong2021science, 10.1093/nar/gkz946} datasets contains 5 different interaction types between genes. 
  (c) Protein-Protein Interaction: illustration of different protein interactions in STRING \cite{szklarczyk2019string}. 
  (d) Single-Cell Type Annotation: t-SNE visualization of zheng68k \cite{zheng2017massively}. Each point indicates a cell embedding and there are 11 cell types. 
  }
    \label{fig4:benchmark}
    \vspace{-10pt}
\end{figure}

\subsubsection{Pathway Co-present Prediction (Gene-level)}

Gene pathway refers to a network of genes and their encoded products that work together to carry out specific biological functions within a cell, which is visualized in Figure \ref{fig4:a}. These pathways illustrate how genes regulate cellular processes through their expression and interactions. Genes produce various proteins and other molecules that participate in a series of regulated steps, leading to cellular responses such as metabolism, signal transduction, and gene regulation\cite{joshi2005reactome}. Accurately predicting pathway co-presence genes requires the embedding to capture complex biological relationships, interactions, and functional similarities between genes, which is a fundamental aspect of gene embedding quality.
The positive and negative gene pairs are formed based on whether a pair of genes is co-present in the same pathway. We provide a clear train test split with non-overlapping genes with each gene occurrence less than 18 times. There are 83,998 positive pairs and 82,796 negative pairs of 4,678 genes in the train set. In the test set, 1,171 genes are used to generate 20,831 positive pairs and 19,972 negative pairs. We perform 5-fold cross-validation and report the average accuracy.

\subsubsection{Functional Gene Interaction Prediction (Gene-level)}

Functional Gene Interaction Prediction is a task aimed at categorizing various types of functional interactions between gene pairs based on their biological roles and regulatory mechanisms, which is illustrated in Figure \ref{fig4:b}. This classification includes interactions where one gene controls the expression, or state change of another, genes that interact or form complexes, and catalytic processes. This task facilitates the elucidation of the intricate web of genetic interactions that govern cellular functions, thereby validating the effectiveness of gene embedding in capturing gene-level biologically relevant information \cite{evans2023transcriptome, wong2021science, 10.1093/nar/gkz946}. 

The source data is obtained from PathwayCommons \cite{wong2021science, 10.1093/nar/gkz946}. After filtering for valid gene interactions, we divided the data into training and testing sets with non-overlapping genes. The training dataset comprises 202,192 gene pairs involving 7,171 unique genes, while the testing dataset contains 50,500 gene pairs with 3,124 unique genes. We employed 5-fold cross-validation to evaluate the task, reporting the mean accuracy obtained.

\subsubsection{Protein-Protein Interaction (Protein-level)}

Protein-protein interactions (PPI) are crucial for understanding the cellular machinery, as they form the backbone of cellular functions and processes, as visualized in Figure \ref{fig4:c}. These interactions encompass both direct physical contacts and indirect functional associations between proteins. The comprehensive mapping and analysis of PPI networks are essential for understanding the complex interplay between proteins that drive biological functions, from metabolic pathways to signal transduction and cellular regulation\cite{szklarczyk2019string}. The biological impact of studying PPIs includes insights into disease mechanisms, identification of biomarkers, and potential therapeutic targets, thus advancing our understanding of cellular processes and developing novel medical interventions. 

The STRING database integrates experimental data, computational predictions, and knowledge from curated databases to provide a global network of protein associations, facilitating functional discovery in genome-wide experimental datasets\cite{szklarczyk2019string}. STRING dataset with valid pairs contains 3,956 proteins and 405,566 PPIs. We follow the Breadth-First Search (BFS) and Depth-First
Search (DFS) methods to split the training and testing PPI datasets\cite{lv2021learning}. Accuracy at different splitting is reported for this PPI task. 

\subsubsection{Single-Cell Type Annotation (Cell-level)}

Single-cell RNA sequencing (scRNA-seq) is crucial for Single-cell type annotation as it allows for the characterization of the transcriptome at the individual cell level,  facilitating the discovery of novel cell types and providing insights into complex regulatory networks during development and disease processes \cite{zheng2017massively}. The expression levels for a set of genes compose the transcriptomic data for cells. The Single-Cell Type Annotation benchmark could evaluate the ability of gene embedding to capture intricate patterns of gene expression and interactions specific to different cell types. 

We select Zheng68k dataset as a representative benchmark for Single-Cell Type Annotation. Zheng68k contains single-cell RNA-seq transcriptomic profiles of 68,450 peripheral blood mononuclear cells (PBMCs), including T cells, B cells, NK cells, monocytes, and dendritic cells, providing a comprehensive overview of immune cell diversity\cite{zheng2017massively}. We plot the t-SNE visualization of it in Figure \ref{fig4:d}. It measures the transcriptomics of 16,906 genes across 11 types of immune cells. We perform 5-fold validation with the setting of scBERT \cite{yang2022scbert} for the Zheng68k benchmark. Accuracy scores are reported as the evaluation metrics for this task. 

\section{Experiments and Results}

\subsection{Experiments Setup}

To evaluate the gene embedding quality, three baselines are selected. Gene2Vec\cite{du2019gene2vec}, an expression-based model, generates gene embedding utilizing gene co-expression information. DNAbert-2\cite{zhou2023dnabert2} is a sequence-based model that generates gene embedding from DNA sequence. OntoProtein\cite{zhang2022ontoprotein} leverage GO knowledge graph in protein-pertaining models to generate protein embedding from amino acid sequences. To show the effectiveness and informativeness of the GOA dataset, we design a simple benchmark gene functional embedding namely GOA\_Emb. LLMs \cite{emb2024mxbai, li2023angle} are used to generate text embedding for each GO term in GO DAG, and then gene embedding is generated by pooling the GOA for each gene. GOA\_Emb represents the simplest gene functional information. As Gene2Vec represents gene expression information, and DNAbert-2 represents the gene sequence information, we derive two more benchmark gene embedding by combining GOA\_Emb with Gene2Vec and DNAbert-2 respectively to show the completion of different aspects of gene information.

\subsection{Results Analysis}
Table \ref{tab:results} records the results of four evaluation benchmarks, where the accuracy score is calculated as the evaluation metric for each task. The results demonstrate the effectiveness of GOA\_Emb and the information completion between different aspects of gene information.  
\begin{table}[!b]
    \centering
    \caption{Results of Benchmark Evaluation Tasks. Accuracy or averaged accuracy with standard deviation is reported. EXP represents the gene-expression-based model, SEQ denotes sequence-based model, and FUN stands for function-based model. "|" separates the results of BFS and DFS.}
    \label{tab:results}
    \resizebox{\textwidth}{!}{%
    \begin{tabular}{lcccccc}
        \toprule
        \multirow{4}{*}{\large \textbf{Model Type}} & \multirow{4}{*}{\large \textbf{Model}}  & \multicolumn{2}{c}{\textbf{Gene-Level}} & \textbf{Protein-Level} & \textbf{Cell-Level} \\
         \cmidrule(lr){3-4} \cmidrule(lr){5-5} \cmidrule(lr){6-6}
     & &  Pathway Co-present & Functional Gene & Protein-Protein & Single-Cell Type\\
     & &  Prediction  & Interaction Prediction & Interaction  & Annotation \\
        \cmidrule(lr){3-4} \cmidrule(lr){5-5} \cmidrule(lr){6-6}
         & & MsigDB & PathwayCommons & STRING (BFS | DFS) & Zheng68k \\
        \midrule
        EXP & Gene2Vec \cite{du2019gene2vec}  & 67.62 $\pm$ 0.39 & 76.16 $\pm$ 1.71 & 79.29 | \textbf{88.19} & 77.86 $\pm$ 0.13 \\
        SEQ & DNABert-2 \cite{zhou2023dnabert2} & 60.97 $\pm$ 0.25 & 58.54 $\pm$ 0.03 & 76.97 | 76.29 & 74.35 $\pm$ 0.35 \\
        SEQ + FUN & OntoProtein \cite{zhang2022ontoprotein} & 48.95 $\pm$ 0.00 & 57.20 $\pm$ 0.00 & 72.33 | 75.43 & 73.58 $\pm$ 0.14 \\
        \midrule
        FUN & GOA\_Emb & 65.33 $\pm$ 0.95 & 75.06 $\pm$ 1.07 & \textbf{88.72} | 87.34 & 77.07 $\pm$ 0.25 \\
        FUN + EXP & GOA\_Emb + Gene2Vec  & \textbf{70.36 $\pm$ 0.21} & \textbf{81.98 $\pm$ 0.71} & 83.18 | 83.05 & \textbf{78.80 $\pm$ 0.20} \\
        FUN + SEQ & GOA\_Emb + DNABert-2 & 68.87 $\pm$ 0.32 & 76.07 $\pm$ 0.41 & 81.49 | 82.13 & 77.07 $\pm$ 0.22 \\
        \bottomrule
    \end{tabular}
    }
\vspace{-10pt}
\end{table}

\begin{table}[!ht]
    \centering
    \caption{Results of the Protein-Protein interaction task on STRING dataset under BFS vs. DFS.}
    \label{tab:resultsppi}
    \resizebox{\textwidth}{!}{%
    \begin{tabular}{@{}lcccccccccc@{}}
        \toprule
        \textbf{Model Type} & \textbf{Model} & \multicolumn{4}{c}{\textbf{BFS}} & \multicolumn{4}{c}{\textbf{DFS}} \\ 
        \cmidrule(r){3-6} \cmidrule(l){7-10}
        & & \textbf{Acc} & \textbf{F1} & \textbf{Recall} & \textbf{Precision} & \textbf{Acc} & \textbf{F1} & \textbf{Recall} & \textbf{Precision} \\ 
        \midrule
        
        EXP & Gene2Vec \cite{du2019gene2vec} & 79.29 & 55.20 & 44.80 & 71.90 & \textbf{88.19} & \textbf{78.80} & 75.63 & \textbf{82.24} \\
        SEQ & DNAbert-2 \cite{zhou2023dnabert2} & 76.97 & 49.34 & 46.59 & 52.45 & 76.29 & 59.36 & 59.22 & 59.50 \\
        SEQ + FUN & OntoProtein \cite{zhang2022ontoprotein}& 72.33 & 48.68 & 54.67 & 43.87 & 75.43 & 60.09 & 62.98 & 57.45 \\
        \midrule
        FUN & GOA\_Emb & \textbf{88.72} & \textbf{80.06} & \textbf{76.29} & \textbf{84.21} & 87.34 & 77.91 & \textbf{83.73} & 72.85 \\
        FUN + EXP & GOA\_Emb + Gene2Vec & 83.18 & 62.48 & 56.16 & 70.39 & 83.05 & 69.33 & 62.74 & 77.46 \\
        FUN + SEQ & GoA\_Emb + DNAbert-2 & 81.49 & 63.17 & 59.49 & 67.35 & 82.13 & 66.49 & 61.07 & 72.96 \\
        \bottomrule
    \end{tabular}
    }
    \vspace{-10pt}
\end{table}

For the gene-level benchmark evaluation tasks, the average accuracy of 5-fold cross-validation is reported in Table \ref{tab:results}. As observed, OntoProtein\cite{zhang2022ontoprotein} as a pre-trained model intake amino acid sequence, failed to capture the gene-level interaction or functional mechanisms, which indicates the existence of an information gap between the protein-pre-trained model and gene-level benchmarks. The performance of both GOA\_Emb and Gene2Vec are ordinary, while the combination of these achieves significant improvement. These results support that the complementary between gene expression and functional information enhances the performance of downstream applications.

For the protein-level evaluation benchmark, PPI on the STRING dataset with accuracy is reported in Table \ref{tab:results}, and detailed results are in Table \ref{tab:resultsppi}.
Gene2Vec and GOA\_Emb achieve the highest performances on BFS and DFS splitting respectively, which reinforce that the gene embedding generated on gene information could capture the essential information or knowledge of their gene products and generalize to the gene products-related applications. 

The results of the single-cell Type annotation evaluation benchmark in Table \ref{tab:results} also reflect the effectiveness of gene information complementary between expression and function aspects, as \textbf{GOA\_Emb + Gene2Vec} achieves the highest accuracy. Moreover, the combination of gene sequence and functional information also improves compared to the results of using single information.

\section{Discussion and Future Work}

Our study provides a template to prove the availability of leveraging ontology and corresponding annotation in real-world applications. There are still challenges that need to be solved. 1) There are more than 2 million species according to NCBI Taxonomy, current work is far from getting all annotations manually reviewed for all the species. Partially because some species are studied but not collected into databases or annotated into GOA. UniEntrezDB will continue unifying more available public databases with GOA. 2) For the four evaluation benchmarks, other similar datasets do not have a unified gene identifier. More datasets can be unified to test gene embedding qualities. 3) This work provides fundamental experiments that prove the effectiveness and informativeness of embedding generated from UniEntrezDB GOA. How to neatly incorporate large-scale GOA data into foundation models to enhance real-world applications is still an open question. 
\section{Conclusion}
We propose a Unified Entrez Gene Identifier Dataset and Benchmarks (UniEntrezDB) to address the challenges of integrating large-scale domain knowledge into gene-related research. This dataset aims to standardize and unify gene ontology annotations from various sources, using unique gene identifiers to create a comprehensive and reliable resource. Future efforts could be conducted to expand and standardize gene databases, improve model training methodologies, and refine the mapping between gene names and identifiers. Incorporating more datasets and exploring alternative annotation methods will also enhance the dataset's robustness.

\newpage
\bibliographystyle{plainnat}
\bibliography{ref}

\end{document}


\appendix

\section{Data Resources and Availability} \label{appendix:resource}
Based on the dataset statistics, this dataset incorporates gene databases with different identifiers from fourteen sources, including 
Xenopus Biology Database (Xenbase)\cite{karimi2018xenbase}, 
Saccharomyces Genome Database (SGD)\cite{cherry2012saccharomyces},  
Zebrafish Information Network (ZFIN)\cite{howe2012zfin}, 
FlyBase (FB)\cite{attrill2016flybase}, 
AmoebaDB \cite{aurrecoechea2010amoebadb}, 
CryptoDB \cite{heiges2006cryptodb},
FungiDB \cite{basenko2024new},
GiardiaDB \cite{aurrecoechea2009giardiadb},
JaponicusDB \cite{rutherford2022japonicusdb},
PlasmoDB \cite{aurrecoechea2009plasmodb},
RefSeq \cite{o2016reference, pruitt2007ncbi},
RNACentral \cite{rnacentral2021rnacentral},
ToxoDB \cite{harb2020toxodb},
TriTryDB \cite{shanmugasundram2023tritrypdb},
Ensembl \cite{martin2023ensembl},
The Arabidopsis Information Resource (TAIR)\cite{berardini2015arabidopsis},  
Mouse Genome Informatics (MGI)\cite{bult2019mouse}, 
PomBase\cite{wood2012pombase}, 
Universal Protein Resource Knowledgebase (UniProtKB)\cite{uniprot2019uniprot}, 
Pseudomonas Genome Database (PseudoCAP)\cite{winsor2016enhanced}, 
Candida Genome Database (CGD)\cite{skrzypek2016candida}, 
and dictyBase\cite{basu2012dictybase}.
We have aligned the gene identifiers from these sources with Gene EntrezID. The Gene EntrezID is derived from Entrez Gene, a comprehensive gene-specific database managed by the National Center for Biotechnology Information (NCBI)\cite{maglott2005entrez}. This repository collects extensive gene-specific information from a broad spectrum of fully sequenced genomes, especially those with active research communities or subject to detailed sequence analysis\cite{maglott2005entrez}. 

This section provides detailed information on the available code and downloadable data from UniEntrezDB. We include links for accessing the ID mapping dictionaries, the UniEntrezGOA dataset, and the benchmark evaluation data. Additionally, we list the sources of the ID mappings and GOA data for each database included in UniEntrezDB. This ensures that users have comprehensive access to the necessary resources for their research and can trace the provenance of the data to its sources, enhancing transparency and reproducibility.


\subsection{Code and Data Availability}

To ensure the reproducibility of experimental results, the source code for this project is hosted on GitHub and can be accessed via the following URL: \url{https://github.com/MM-YY-WW/UniEntrezDB.git}. This repository includes the gene embeddings for the six benchmark methods, as well as the gene EntrezIDs, DNA sequences, and amino acid sequences involved in all datasets. Additionally, it contains the model checkpoints corresponding to the reported experimental results, allowing researchers to replicate the results.

We provide the ID mapping relationships among 21 databases. These curated ID mappings are available in JSON format, facilitating seamless integration and analysis. Compressed versions of the ID mapping dictionaries, totaling 858.4 MB, are available for public download at \url{https://drive.google.com/file/d/1La80B3hUibbe94FghkTIx80DRzPfwYix/view?usp=sharing}.

Our UniEntrezGOA dataset includes the 11 columns required for GAF format files along with an additional EntrezID column. This ensures comprehensive annotation and easy cross-referencing with other databases. The compressed dataset for UniEntrezGOA, which is 661.8 MB in size, is accessible at \url{https://drive.google.com/file/d/1DsXufybeSgEXrx8szkF0kuhASmAVOaU-/view?usp=sharing}.

The UniEntrez Benchmark Evaluations, which include four downstream tasks standardized with EntrezID and data splits, are also available for download. These evaluations facilitate the assessment of gene function prediction, interaction prediction, and other bioinformatics tasks. The compressed size of this dataset is 300.7 MB and can be accessed at \url{https://drive.google.com/file/d/1fSRXO26jr1XcFn7GKqRoN_CZUbuEY8Cj/view?usp=sharing}.

\subsection{Data Source of Unifying Gene EntrezID} 
For the UniEntrezID GOA Dataset, we mainly perform the Gene EntrezID unifying based on the ID mapping files on the File Transfer Protocol (FTP) site provided by specialized databases. This section introduces the source ID mapping files for each database. 

\newpage

\paragraph{UniProtKB}
The ID mapping source file from UniProtKB\cite{uniprot2019uniprot} to Gene EntrezID\cite{maglott2005entrez} is available at \url{https://ftp.uniprot.org/pub/databases/uniprot/current_release/knowledgebase/idmapping/idmapping_selected.tab.gz}. In this process, 13,467,343 ids are successfully mapped to EntrezID by UniProtKB.

\paragraph{Ensembl}
The mapping between Ensembl\cite{martin2023ensembl} and Entrez Gene \cite{maglott2005entrez} is \url{https://ftp.ncbi.nih.gov/gene/DATA/gene2ensembl.gz}. Ensembl has accomplished the mapping of 35,535,857 IDs to EntrezID.

\paragraph{RefSeq}
The RefSeq\cite{o2016reference, pruitt2007ncbi} ID are mapped to Entrez Gene ID\cite{maglott2005entrez} via \url{https://ftp.ncbi.nih.gov/gene/DATA/gene2refseq.gz}. Through this resource, RefSeq has mapped 129,642,994 IDs to EntrezID.

\paragraph{RNACentral}
RNACentral\cite{rnacentral2021rnacentral} provides their ID mapping file at \url{https://ftp.ebi.ac.uk/pub/databases/RNAcentral/releases/24.0/id_mapping/database_mappings/ensembl.tsv} and \url{https://ftp.ebi.ac.uk/pub/databases/RNAcentral/releases/24.0/id_mapping/database_mappings/refseq.tsv} for mapping to Ensembl\cite{martin2023ensembl} and RefSeq\cite{o2016reference, pruitt2007ncbi}, respectively. RNACentral mapping 198,009 IDs with EntrezID.

\paragraph{MGI}
MGI\cite{bult2019mouse} map its ID to UniProtKB\cite{uniprot2019uniprot}, RefSeq\cite{o2016reference, pruitt2007ncbi}, and Ensembl\cite{martin2023ensembl} by \url{https://www.informatics.jax.org/downloads/reports/gp2protein.mgi}. MGI has a total of 17,508 IDs linked to EntrezID.

\paragraph{ZFIN}
ZFIN\cite{howe2012zfin} use \url{https://zfin.org/downloads/ensembl_1_to_1.txt} to map id to Ensembl\cite{martin2023ensembl} and \url{https://zfin.org/downloads/uniprot-zfinpub.txt} to map id to UniProtKB\cite{uniprot2019uniprot}. From the ZFIN database, 21,379 IDs have been mapped to EntrezID.

\paragraph{CGD}
CGD ID maps to Entrez Gene ID by \url{http://www.candidagenome.org/download/External_id_mappings/CGDID_2_GeneID.tab.gz} and to UniProtKB ID by \url{http://www.candidagenome.org/download/External_id_mappings/gp2protein.cgd.gz}. CGD contributed to the mapping of 24,298 IDs to EntrezID.

\paragraph{PseudoCAP}
PseudoCAP\cite{winsor2016enhanced} have 1473 ids mapped successfully to UniProtKB\cite{uniprot2019uniprot} with \url{https://ftp.ebi.ac.uk/pub/databases/GO/goa/gp2protein/gp2protein.pseudocap.gz}. A total of 1,473 IDs have been effectively mapped to EntrezID by PseudoCAP.

\paragraph{TriTrypDB}
TriTrypDB\cite{shanmugasundram2023tritrypdb} provide mapping files at \url{https://tritrypdb.org/tritrypdb/app/downloads} to map id to UniProtKB\cite{uniprot2019uniprot}. TriTrypDB has the mapping of 46,533 IDs to EntrezID.

\paragraph{dictybase}
dictybase\cite{basu2012dictybase} use \url{http://dictybase.org/db/cgi-bin/dictyBase/download/download.pl?area=general&ID=DDB-GeneID-UniProt.txt} to map id to UniProtKB\cite{uniprot2019uniprot}. The dictybase maps 12,189 IDs to EntrezID.

\paragraph{CryptoDB}
CryptoDB\cite{heiges2006cryptodb} provides a download portal for mapping files at \url{https://cryptodb.org/cryptodb/app/downloads} for mapping to UniProtKB\cite{uniprot2019uniprot}. From the resources of CryptoDB, 7,740 IDs have now been linked to EntrezID.

\paragraph{PomBase}
PomBase\cite{wood2012pombase} maps id to UniProtKB id by \url{https://www.pombase.org/data/names_and_identifiers/PomBase2UniProt.tsv}. PomBase show 4,644 IDs linked to EntrezID.

\paragraph{FB}
ID mapping file of FB\cite{attrill2016flybase} to UniProtKB\cite{uniprot2019uniprot} ID is \url{http://ftp.flybase.org/releases/FB2024_01/precomputed_files/genes/fbgn_NAseq_Uniprot_fb_2024_01.tsv.gz}. FB indicates 11,322 ids successfully mapped to EntrezID.

\paragraph{TAIR}
ID mapping file of TAIR\cite{berardini2015arabidopsis} database to UniProtKB\cite{uniprot2019uniprot} is \url{https://www.arabidopsis.org/download_files/Proteins/Id_conversions/TAIR2UniprotMapping.txt}. This mapping effort has linked 72,103 IDs from TAIR to EntrezID.

\paragraph{FungiDB}
The download portal of FungiDB\cite{basenko2024new} at \url{https://fungidb.org/fungidb/app/downloads} offers a mapping file to UniProtKB\cite{uniprot2019uniprot} ID, resulting in 462,853 IDs being mapped to EntrezID.

\paragraph{GiardiaDB}
GiardiaDB\cite{aurrecoechea2009giardiadb} also provides the download portal for mapping files to UniProtKB\cite{uniprot2019uniprot} at \url{https://giardiadb.org/giardiadb/app/downloads}. GiardiaDB mapping process includes 4,032 IDs that are now associated with EntrezID.

\paragraph{Xenbase}
Xenbase\cite{karimi2018xenbase} provide UniProtKB\cite{uniprot2019uniprot} id mapping at \url{https://download.xenbase.org/xenbase/DataExchange/Uniprot/XenbaseGeneUniprotMapping.txt}. Xenbase has mapped 31,581 IDs to EntrezID effectively.

\paragraph{AmoebaDB}
ID mapping from AmoebaDB\cite{aurrecoechea2010amoebadb} to UniProtKB can be downloaded at \url{https://amoebadb.org/amoeba/app/downloads}. AmoebaDB has facilitated the mapping of 46,770 IDs to EntrezID.

\paragraph{JaponicusDB}
JaponicusDB\cite{rutherford2022japonicusdb} maps to UniProtKB\cite{uniprot2019uniprot} by \url{https://www.japonicusdb.org/data/names_and_identifiers/JaponicusDB2UniProt.tsv}. The database has 4,878 IDs being mapped to EntrezID.

\paragraph{ToxoDB}
ID mapping files to UniProtKB\cite{uniprot2019uniprot} of ToxoDB\cite{harb2020toxodb} can be obtained via download portal \url{https://toxodb.org/toxo/app/downloads}. ToxoDB achieved the mapping of 8,644 IDs to EntrezID.

\paragraph{SGD}
SGD\cite{cherry2012saccharomyces} maps its id to UniProtKB\cite{uniprot2019uniprot} by \url{https://sgd-prod-upload.s3.amazonaws.com/S000214964/dbxref.20170114.tab.gz}. SGD accounted for the mapping of 5,909 IDs to EntrezID.

\paragraph{PlasmoDB}
Finally, PlasmoDB\cite{aurrecoechea2009plasmodb} provide download site for mapping files to UniProtKB\cite{uniprot2019uniprot} at \url{https://plasmodb.org/plasmo/app/downloads}. PlasmoDB has successfully mapped 23,837 IDs to EntrezID.

\subsubsection{Data Source of GOA}
Table \ref{appendixtab:database-stats} provides an overview of statistics of GOA obtained from 21 databases. This section will provide the source data links for each database.

\paragraph{UniProtKB}
GOA files for UniProtKB\cite{uniprot2019uniprot} are accessible at \url{https://ftp.ebi.ac.uk/pub/databases/GO/goa/proteomes}.

\paragraph{RefSeq}
GOA resources for RefSeq\cite{o2016reference, pruitt2007ncbi} are available at \url{https://ftp.ncbi.nlm.nih.gov/genomes/refseq/vertebrate_mammalian/Acinonyx_jubatus/annotation_releases/current/GCF_027475565.1-RS_2023_04/GCF_027475565.1-RS_2023_04_gene_ontology.gaf.gz}.

\paragraph{RNACentral}
RNACentral\cite{rnacentral2021rnacentral} provides GOA files at \url{https://ftp.ebi.ac.uk/pub/databases/GO/goa}.

\paragraph{MGI}
Mouse Genome Informatics (MGI)\cite{bult2019mouse} offers GOA data at \url{https://current.geneontology.org/annotations/mgi.gaf.gz}.

\paragraph{ZFIN}
The Zebrafish Information Network (ZFIN)\cite{howe2012zfin} hosts its GOA files at \url{https://current.geneontology.org/annotations/zfin.gaf.gz}.

\paragraph{CGD}
The Candida Genome Database (CGD)\cite{skrzypek2016candida} makes its GOA files available at \url{http://www.candidagenome.org/download/go}.

\paragraph{PseudoCAP}
GOA files for PseudoCAP\cite{winsor2016enhanced} can be found at \url{https://current.geneontology.org/annotations/pseudocap.gaf.gz}.

\paragraph{TriTrypDB}
TriTrypDB\cite{shanmugasundram2023tritrypdb} provides access to its GOA at \url{https://tritrypdb.org/common/downloads/Current_Release}.

\paragraph{dictybase}
GOA files from dictybase\cite{basu2012dictybase} are available for download at \url{http://viewvc.geneontology.org/viewvc/GO-SVN/trunk/gene-associations/submission/gene_association.dictyBase.gz?rev=HEAD}.

\paragraph{CryptoDB}
CryptoDB\cite{heiges2006cryptodb} hosts its GOA resources at \url{https://cryptodb.org/cryptodb/app/downloads}.

\paragraph{PomBase}
PomBase\cite{wood2012pombase} offers GOA files at \url{https://www.pombase.org/data/annotations/Gene_ontology}.

\paragraph{FB}
FlyBase (FB)\cite{attrill2016flybase} provides its GOA data at \url{https://current.geneontology.org/annotations/fb.gaf.gz}.

\paragraph{TAIR}
The Arabidopsis Information Resource (TAIR)\cite{berardini2015arabidopsis} hosts GOA files at \url{https://www.arabidopsis.org/download/file?path=GO_and_PO_Annotations/Gene_Ontology_Annotations/gene_association.tair.gz}.

\paragraph{FungiDB}
FungiDB\cite{basenko2024new} makes its GOA available at \url{https://fungidb.org/fungidb/app/downloads}.

\paragraph{GiardiaDB}
GiardiaDB\cite{aurrecoechea2009giardiadb} provides access to GOA files at \url{https://giardiadb.org/giardiadb/app/downloads}.

\paragraph{Xenbase}
Xenbase\cite{karimi2018xenbase} offers GOA files at \url{https://current.geneontology.org/annotations/xenbase.gaf.gz}.

\paragraph{AmoebaDB}
GOA files for AmoebaDB\cite{aurrecoechea2010amoebadb} are available at \url{https://amoebadb.org/amoeba/app/downloads}.

\paragraph{JaponicusDB}
JaponicusDB\cite{rutherford2022japonicusdb} hosts its GOA at \url{https://www.japonicusdb.org/data/annotations/Gene_ontology}.

\paragraph{ToxoDB}
ToxoDB\cite{harb2020toxodb} provides its GOA files at \url{https://toxodb.org/toxo/app/downloads}.

\paragraph{SGD}
The Saccharomyces Genome Database (SGD)\cite{cherry2012saccharomyces} offers GOA at \url{http://sgd-archive.yeastgenome.org/curation/literature/gene_association.sgd.gaf.gz}.

\paragraph{PlasmoDB}
PlasmoDB\cite{aurrecoechea2009plasmodb} makes its GOA resources available at \url{https://plasmodb.org/plasmo/app/downloads}.

\section{Statistics} \label{appendix:statistics}

Table \ref{appendixtab:database-stats} provides detailed statistics of the GOA data compiled from 21 different databases. Each column in the table represents a specific aspect of the data, offering insights into the composition and mapping efficiency of the annotations. Table \ref{appendixtab:goa_columns} provides a comprehensive description of the columns in the UniEntrezGOA dataset. This table details the total number of entries for each column across all GOA categories and the number of entries excluding those with "IEA" (Inferred from Electronic Annotation) evidence codes.

In Table \ref{appendixtab:taxon}, we present a comprehensive overview of the top 50 taxon IDs from the UniEntrezGOA dataset, excluding entries with IEA (Inferred from Electronic Annotation) evidence codes. This table lists the corresponding taxon names and the number of Gene Ontology Annotations (GOA) for each taxon. The purpose of this table is to illustrate the diversity and richness of species represented in the UniEntrezGOA dataset. For instance, Schizosaccharomyces pombe 972h- (taxon:284812) tops the list with 821,443 GO annotations, followed by Mus musculus (taxon:10090) with 658,139 annotations and Aspergillus fumigatus Af293 (taxon:330879) with 634,536 annotations. This diversity spans a wide range of organisms, including model organisms like Homo sapiens (taxon:9606) with 423,865 annotations, and Arabidopsis thaliana (taxon:3702) with 130,727 annotations. 
We visualize the complete version of the phylogenetic tree formed by the species without IEA GOA at the end.

Table \ref{tab:addlabel} presents the counts and percentages of GOA by year, comparing the growth in the number of GOA before and after the introduction of IEA (Inferred from Electronic Annotation) methods. This table highlights the annual increase in GOA and emphasizes the shift from manually curated annotations to the use of automated IEA methods. Before 2017, all GOA relied on manual review and annotation, as evidenced by the relatively steady and moderate increases in the number of annotations each year.

From 2017 onwards, there has been a notable surge in the number of GOA, coinciding with the widespread adoption of IEA methods. For example, in 2017, the number of GOA without IEA was 782,158, while the total number of GOA, including IEA, jumped to 7,706,748. This trend continues, with significant increases in the number of GOA each subsequent year, peaking in 2024 with a total of 44,434,156 GOA, 468,726 of which were manually annotated without IEA.

This upward trend in the number of GOA demonstrates the rapid growth and expansion of the UniEntrezDB dataset, underscoring its potential for providing comprehensive and up-to-date biological annotations. The continual expansion of GOA reflects the increasing integration of automated annotation methods, which enhance the scalability and efficiency of GOA generation. Looking forward, we are committed to regularly updating UniEntrezDB to include more recent GOA, thereby ensuring its relevance and utility for the scientific community.

Table \ref{appendixtab:DB_object_Type} summarizes the counts and percentages of different database (DB) object types in the GOA, divided into proteins, genes, RNA, and other gene products. Proteins are the most represented category, with 47,384,341 annotations (59.10\%) and 3,664,335 without IEA (49.13\%). The gene category, primarily "protein\_coding," has 11,660,397 annotations (14.54\%) and 2,262,415 without IEA (30.33\%). The RNA category, mainly "transcript," includes 20,150,004 annotations (25.13\%) and 939,660 without IEA (12.60\%). Other types include "gene\_product," with 163,226 annotations (0.20\%) and 61,758 without IEA (0.83\%).

Table \ref{tab:goa_distribution} lists the evidence codes, their full names, categories, and their proportions within the 7,458,326 GOA excluding IEA. The table reveals the distribution of GOA across different evidence types, highlighting the data's diversity. Author statements include TAS and NAS, accounting for 2.01\% and 0.66\% of GOA, respectively. Curator statements, such as ND and IC, make up 13.25\% and 0.62\%. Computational analysis evidence codes like ISS and ISO represent 6.44\% and 6.35\%, respectively. Experimental evidence, including IDA (9.23\%), IMP (7.95\%), and IPI (2.98\%), demonstrates significant contributions. Phylogenetically-inferred annotations, mainly IBA, form the largest category at 41.84\%.

\begin{table}[htbp]
    \centering
    \caption{Statistics of the GOA data from 21 databases. The ``GOA with mapped id" column indicates the total number of GOA mapped to EntrezID successfully for each database. The ``\% of GOA with mapped id" indicates the percentage contribution of each database to the total number of GOA with mapped id. ``All GOA from database" is the total number of GOA obtained from each database. ``\% of All GOA from database" indicates the percentage contribution of all GOA from each database to the total number of All GOA. ``\% of GOA with mapped id/ All" indicates the percentage of GOA successfully mapped to the EntrezID for each database. ``GOA not IEA" is the number of GOA mapped to EntrezID and the Evidence Code is not ``IEA". After removing duplication GOA between databases, ``UniEntrezGOA" row indicates there are 80,179,706 valid GOA in resulting dataset in which 7,458,326 GOA do not have a ``IEA" Evidence Code}
    \label{appendixtab:database-stats}
    \resizebox{\textwidth}{!}{%
    \begin{tabular}{l|rrrrrr}
        \toprule
        Database & GOA with & \% of GOA & All GOA  & \% of All GOA  & \% of GOA with & GOA not \\
        
        name & mapped id &with mapped id & from database & from database & mapped id/All & IEA \\
        
        \midrule
        AmoebaDB\cite{aurrecoechea2010amoebadb} & 4,378,744 & 2.81\% & 10,471,510 & 0.98\% & 41.82\% & 255 \\
        CGD\cite{skrzypek2016candida} & 601,769 & 0.39\% & 1,376,480 & 0.13\% & 43.72\% & 155,684 \\
        CryptoDB\cite{heiges2006cryptodb} & 926,666 & 0.59\% & 7,924,600 & 0.74\% & 11.69\% & 68 \\
        dictybase\cite{basu2012dictybase} & 73,525 & 0.05\% & 74,526 & 0.01\% & 98.66\% & 39,036 \\
        FB\cite{attrill2016flybase} & 104679 & 0.07\% & 133,742 & 0.01\% & 78.27\% & 88,688 \\
        FungiDB\cite{basenko2024new} & 58,500,934 & 37.49\% & 186,878,409 & 17.51\% & 31.30\% & 6,358,893 \\
        GiardiaDB\cite{aurrecoechea2009giardiadb} & 443,158 & 0.28\% & 2,574,255 & 0.24\% & 17.22\% & 126,291 \\
        JaponicusDB\cite{rutherford2022japonicusdb} & 33,868 & 0.02\% & 34,000 & 0.00\% & 99.61\% & 2,889 \\
        MGI\cite{bult2019mouse} & 420,521 & 0.27\% & 493,961 & 0.05\% & 85.13\% & 354,256 \\
        PlasmoDB\cite{aurrecoechea2009plasmodb} & 87,164 & 0.06\% & 605,814 & 0.06\% & 14.39\% & 32,697 \\
        pombase\cite{wood2012pombase} & 36,782 & 0.02\% & 41,047 & 0.00\% & 89.61\% & 35,054 \\
        PseudoCAP\cite{winsor2016enhanced} & 3,354 & 0.00\% & 3,446 & 0.00\% & 97.33\% & 3,354 \\
        refseq\cite{o2016reference, pruitt2007ncbi} & 85,920 & 0.06\% & 85,920 & 0.01\% & 100.00\% & 0 \\
        RNACentral\cite{rnacentral2021rnacentral} & 43,237,600 & 27.71\% & 398,339,870 & 37.33\% & 10.85\% & 5,711,350 \\
        SGD\cite{cherry2012saccharomyces} & 145,113 & 0.09\% & 150,960 & 0.01\% & 96.13\% & 64,090 \\
        TAIR\cite{berardini2015arabidopsis} & 36,490 & 0.02\% & 280,180 & 0.03\% & 13.02\% & 36,490 \\
        ToxoDB\cite{harb2020toxodb} & 758307 & 0.49\% & 22,158,068 & 2.08\% & 3.42\% & 6,331 \\
        TriTryDB\cite{shanmugasundram2023tritrypdb} & 234,742 & 0.15\% & 1,376,076 & 0.13\% & 17.06\% & 181,031 \\
        uniprot\cite{uniprot2019uniprot} & 4,5480,697 & 29.14\% & 433,621,091 & 40.63\% & 10.49\% & 3,564,408 \\
        Xenbase\cite{karimi2018xenbase} & 263,047 & 0.17\% & 344,228 & 0.03\% & 76.42\% & 40,953 \\
        ZFIN\cite{howe2012zfin} & 202,316 & 0.13\% & 222,765 & 0.02\% & 90.82\% & 100,957 \\
        \midrule
        Total & 156,055,396 & 100.00\% & 1,067,190,948 & 100.00\% & 24.63\% & 16,902,775 \\
        UniEntrezGOA &80,179,706& - & - & - & - & 7,458,326\\
        \bottomrule
    \end{tabular}
    }
\end{table}

\begin{table}[htbp]
  \centering
  \caption{Description and Distribution of Columns in UniEntrezGOA Dataset}
    \resizebox{\textwidth}{!}{%
    \begin{tabular}{cllll}
    \toprule
    \textbf{Column} & \textbf{Column} & \textbf{\# (all GOA } & \textbf{\# (w/o IEA GOA} & \textbf{Descriptions}\\
    \textbf{Number} & \textbf{Name} & \textbf{Categories)} & \textbf{Categories)} & \\
    \midrule
    & & &  & Database source  for the identifier specified in DB\\
    1 & DB & 17 & 16 &  Object ID \\
    \midrule
     &  &  &  & Unique identifier  within the database specified in\\
    2 & DB\_Object\_ID & 6,202,996 & 834,337 &  the 'DB' column \\

    \midrule    
    & & &  &  Unique and valid symbol corresponding to the DB\\
    3 & DB\_Object\_Symbol & 4,051,233 & 586,865 &  Object ID \\
    \midrule    
    4 & GO\_ID & 31,721 & 28,668 & GO identifier associated with the DB Object ID \\

    \midrule    
    
    & & &  &  Identifier(s) of  the source(s) cited as evidence  for \\
    5 & DB:Reference & 657,979 & 257,189 &the GO annotation \\
    \midrule    
    
    & & &  & Code indicating the evidence supporting the GO \\
    6 & Evidence\_Code & 24 & 23 &  annotation \\
    \midrule    
    & & &  & Namespace of the GO ID, indicating if it belongs \\
    & & &  & to Biological Process (P), Molecular Function (F), \\
    7 & Aspect & 3 & 3 &   or Cellular Component (C)   \\
    \midrule    
    & & &  &  Description of the type of entity (gene product) \\
    8 & DB\_Object\_Type & 11 & 10 & being annotated \\
    \midrule    
    & & &  & Taxonomic identifier indicating the organism of the \\
    9 & Taxon & 10,548 & 1,012 & gene product \\
    \midrule    
    10 & Date & 7,631 & 7,628 & Date of annotation in YYYYMMDD format \\
    \midrule    
    11 & Assigned\_By & 84 & 69 & Database responsible for the annotation \\
    \midrule    
    12 & EntrezID & 5,205,671 & 678,634 & Entrez Gene ID corresponding to the DB Object ID \\
    \bottomrule
    \end{tabular}%
    }
  \label{appendixtab:goa_columns}%
\end{table}%

\begin{table}[htbp]
  \centering
  \caption{Top 50 Taxon ID in GOA without IEA evidence code and corresponding Taxon Name, and Number of GOA}
    \resizebox{\textwidth}{!}{%
    \begin{tabular}{llr}
    \toprule
    \textbf{Taxon} & \textbf{Taxon Name} & \textbf{Number of GOA (No IEA)} \\
    \midrule
    taxon:284812 & Schizosaccharomyces pombe 972h- & 821,443 \\
    taxon:10090 & Mus musculus & 658,139 \\
    taxon:330879 & Aspergillus fumigatus Af293 & 634,536 \\
    taxon:367110 & Neurospora crassa OR74A & 582,777 \\
    taxon:9606 & Homo sapiens & 423,865 \\
    taxon:227321 & Aspergillus nidulans FGSC A4 & 337,608 \\
    taxon:10116 & Rattus norvegicus & 278,989 \\
    taxon:7227 & Drosophila melanogaster & 237,620 \\
    taxon:7955 & Danio rerio & 235,082 \\
    taxon:237561 & Candida albicans SC5314 & 156,351 \\
    taxon:559292 & Saccharomyces cerevisiae S288C & 147,576 \\
    taxon:418459 & Puccinia graminis f. sp. tritici CRL 75-36-700-3 & 142,003 \\
    taxon:425011 & Aspergillus niger CBS 513.88 & 139,100 \\
    taxon:3702 & Arabidopsis thaliana & 130,727 \\
    taxon:237631 & Ustilago maydis 521 & 130,238 \\
    taxon:185431 & Trypanosoma brucei brucei TREU927 & 129,168 \\
    taxon:214684 & Cryptococcus neoformans var. neoformans JEC21 & 115,543 \\
    taxon:510516 & Aspergillus oryzae RIB40 & 106,240 \\
    taxon:235443 & Cryptococcus neoformans var. grubii H99 & 73,780 \\
    taxon:4097 & Nicotiana tabacum & 71,214 \\
    taxon:44689 & Dictyostelium discoideum & 70,872 \\
    taxon:3635 & Gossypium hirsutum & 66,699 \\
    taxon:9913 & Bos taurus & 63,204 \\
    taxon:6239 & Caenorhabditis elegans & 61,357 \\
    taxon:184922 & Giardia lamblia ATCC 50803 & 52,460 \\
    taxon:3847 & Glycine max & 46327 \\
    taxon:5888 & Paramecium tetraurelia & 45,906 \\
    taxon:9598 & Pan troglodytes & 43,637 \\
    taxon:5476 & Candida albicans & 35,759 \\
    taxon:4896 & Schizosaccharomyces pombe & 35,054 \\
    taxon:8364 & Xenopus tropicalis & 34,695 \\
    taxon:36329 & Plasmodium falciparum 3D7 & 33,717 \\
    taxon:9595 & Gorilla gorilla gorilla & 32,508 \\
    taxon:4432 & Nelumbo nucifera & 32,381 \\
    taxon:6412 & Helobdella robusta & 30,813 \\
    taxon:3880 & Medicago truncatula & 30,675 \\
    taxon:83333 & Escherichia coli K-12 & 29,055 \\
    taxon:13616 & Monodelphis domestica & 28,419 \\
    taxon:412133 & Trichomonas vaginalis G3 & 27,999 \\
    taxon:39947 & Oryza sativa Japonica Group & 27,368 \\
    taxon:8090 & Oryzias latipes & 27,070 \\
    taxon:9544 & Macaca mulatta & 27,022 \\
    taxon:15368 & Brachypodium distachyon & 26,780 \\
    taxon:51240 & Juglans regia & 26,125 \\
    taxon:3711 & Brassica rapa & 25,073 \\
    taxon:9031 & Gallus gallus & 24,831 \\
    taxon:28377 & Anolis carolinensis & 24,452 \\
    taxon:8355 & Anolis carolinensis & 22,913 \\
    taxon:9823 & Sus scrofa & 22,264 \\
    taxon:10228 & Trichoplax adhaerens & 22,155 \\
    \bottomrule
    \end{tabular}%
    }
  \label{appendixtab:taxon}%
\end{table}%

\begin{table}[htbp]
  \centering
  \caption{Counts and Percentages of GOA by Year}
    \resizebox{\textwidth}{!}{%
    \begin{tabular}{c|c|rrrr}
    \toprule
    IEA exist or not & \textbf{Year} & \textbf{\# (GOA)} & \textbf{\% (GOA)} & \textbf{\# (GOA w/o IEA)} & \textbf{\% (GOA w/o IEA)} \\
    \midrule
    &1999 & 66 & 0.00\% & 66 & 0.00\% \\
    &2000 & 2,573 & 0.00\% & 2,573 & 0.03\% \\
    &2001 & 6,849 & 0.01\% & 6,849 & 0.09\% \\
    &2002 & 11,614 & 0.01\% & 11,614 & 0.16\% \\
    &2003 & 28,847 & 0.04\% & 28,847 & 0.39\% \\
    &2004 & 32,637 & 0.04\% & 32,637 & 0.44\% \\
    No IEA&2005 & 53,603 & 0.07\% & 53,603 & 0.72\% \\
    Only &2006 & 92,427 & 0.12\% & 92,427 & 1.24\% \\
    Manually &2007 & 64,464 & 0.08\% & 64,464 & 0.86\% \\
    Annotation &2008 & 82,203 & 0.10\% & 82,203 & 1.10\% \\
    &2009 & 101,971 & 0.13\% & 101,971 & 1.37\% \\
    &2010 & 96,805 & 0.12\% & 96,805 & 1.30\% \\
    &2011 & 117,983 & 0.15\% & 117,983 & 1.58\% \\
    &2012 & 90,767 & 0.11\% & 90,767 & 1.22\% \\
    &2013 & 156,048 & 0.19\% & 156,048 & 2.09\% \\
    &2014 & 158,502 & 0.20\% & 158,502 & 2.13\% \\
    &2015 & 94,606 & 0.12\% & 94,606 & 1.27\% \\
    &2016 & 89,647 & 0.11\% & 89,647 & 1.20\% \\
    \midrule
    &2017 & 7,706,748 & 9.61\% & 782,158 & 10.49\% \\
    &2018 & 8,961,562 & 11.18\% & 435,593 & 5.84\% \\
    &2019 & 2,601,278 & 3.24\% & 412,618 & 5.53\% \\
    IEA &2020 & 2,275,754 & 2.84\% & 596,791 & 8.00\% \\
    Methods &2021 & 4,422,390 & 5.52\% & 1,147,160 & 15.38\% \\
    Developed &2022 & 3,292,148 & 4.11\% & 804,921 & 10.79\% \\
    &2023 & 5,204,058 & 6.49\% & 1,528,747 & 20.50\% \\
    &2024 & 44,434,156 & 55.42\% & 468,726 & 6.28\% \\
    \bottomrule
    \end{tabular}%
    }
  \label{tab:addlabel}%
\end{table}%

\section{Extra Results}

\paragraph{Pathway Co-present Prediction}
In this section, we present the results of the Pathway Co-present Prediction task using the MsigDB dataset \cite{liberzon2011molecular}. Table \ref{appendixtab:extra_result_copresent} shows the 5-fold cross-validation results for different models, including their accuracy and F1 scores.
\paragraph{Functional Gene Interaction Prediction}
The results for Functional Gene Interaction Prediction using the PathwayCommons dataset \cite{wong2021science, 10.1093/nar/gkz946} are shown in Table \ref{appendixtab:extra_result_GGI}. The table details the performance of various models over 5-fold cross-validation, displaying both accuracy and F1 scores.
\paragraph{Single-Cell Type Annotation}
Table \ref{appendixtab:extra_results_zheng68k} presents the 5-fold cross-validation results for Single-Cell Type Annotation using the Zheng68k dataset \cite{zheng2017massively}. The results show the performance of various models in terms of accuracy and F1 scores.

        
\begin{table}[!ht]
    \centering
    \caption{Extra 5-fold results of Functional Gene Interaction Prediction (PathwayCommons\cite{wong2021science, 10.1093/nar/gkz946})}
    \label{appendixtab:extra_result_GGI}
    \resizebox{\textwidth}{!}{%
    \begin{tabular}{@{}llccccccc|ccccccc@{}}
        \toprule
        \textbf{Model Type} & \textbf{Model} & \multicolumn{6}{c}{\textbf{Accuracy (\%)}} & \multicolumn{8}{c}{\textbf{F1 Score (\%)}} \\ 
        \cmidrule(lr){3-9} \cmidrule(l){10-16}
        & & \textbf{Fold0} & \textbf{Fold1} & \textbf{Fold2} & \textbf{Fold3} & \textbf{Fold4} & \textbf{Avg} & \textbf{Std} & \textbf{Fold0} & \textbf{Fold1} & \textbf{Fold2} & \textbf{Fold3} & \textbf{Fold4} & \textbf{Avg} & \textbf{Std} \\ 
        \midrule
        EXP & Gene2Vec \cite{du2019gene2vec} & 77.49 & 75.59 & 76.38 & 73.55 & 77.81 & 76.16 & 1.71 & 59.21 & 55.17 & 56.44 & 46.00 & 59.63 & 54.31 & 5.52 \\

        SEQ  & DNAbert-2 \cite{zhou2023dnabert2} & 58.51 & 58.53 & 58.58 & 58.53 & 58.56 & 58.54 & 0.03 & 21.47 & 21.60 & 21.40 & 21.00 & 21.51 & 21.38 & 0.23 \\
        SEQ + FUN & Ontoprotein \cite{zhang2022ontoprotein}& 57.20 & 57.20 & 57.20 & 57.20 & 57.20 & 57.20 & 0.00 & 14.55 & 14.55 & 14.55 & 14.55 & 14.55 & 14.55 & 0.00 \\
        \midrule
        FUN & GOA\_Emb & 73.94 & 76.76 & 75.05 & 74.43 & 75.13 & 75.06 & 1.07 & 57.73 & 67.23 & 60.67 & 58.82 & 59.27 & 61.50 & 3.78 \\
            FUN + EXP & GOA\_Emb + Gene2Vec & \textbf{82.58} & \textbf{82.62} & \textbf{81.97} & \textbf{81.86} & \textbf{80.86} & \textbf{81.98} & 0.71 & \textbf{74.34} & \textbf{74.37} & \textbf{73.44} & \textbf{72.95} & \textbf{71.18} & \textbf{72.99} & 1.31 \\
        FUN + SEQ & GoA\_Emb + DNAbert-2 & 76.26 & 75.81 & 76.70 & 75.75 & 75.83 & 76.07 & 0.41 & 60.98 & 60.08 & 61.19 & 60.29 & 63.75 & 61.33 & 1.47 \\
        \bottomrule
    \end{tabular}
    }
\end{table}

\begin{table}[!ht]
    \centering
    \caption{Extra 5-fold results of Pathway Co-present Prediction (MsigDB\cite{liberzon2011molecular})}
    \label{appendixtab:extra_result_copresent}
    \resizebox{\textwidth}{!}{%
    \begin{tabular}{@{}llccccccc|ccccccc@{}}
        \toprule
        \textbf{Model Type} & \textbf{Model} & \multicolumn{6}{c}{\textbf{Accuracy (\%)}} & \multicolumn{8}{c}{\textbf{F1 Score (\%)}} \\ 
        \cmidrule(lr){3-9} \cmidrule(l){10-16}
        & & \textbf{Fold0} & \textbf{Fold1} & \textbf{Fold2} & \textbf{Fold3} & \textbf{Fold4} & \textbf{Avg} & \textbf{Std} & \textbf{Fold0} & \textbf{Fold1} & \textbf{Fold2} & \textbf{Fold3} & \textbf{Fold4} & \textbf{Avg} & \textbf{Std} \\ 
        \midrule
        EXP & Gene2Vec \cite{du2019gene2vec} & 67.34 & 67.91 & 67.87 & 67.93 & 67.07 & 67.62 & 0.39 & 67.34 & 67.91 & 67.87 & 67.92 & 67.02 & 67.68 & 0.41 \\
        SEQ & DNAbert-2 \cite{zhou2023dnabert2} & 61.11 & 60.87 & 60.62 & 60.97 & 61.28 & 60.97 & 0.25 & 60.81 & 60.50 & 60.48 & 60.81 & 61.00 & 60.70 & 0.22 \\
        SEQ + FUN & OntoProtein \cite{zhang2022ontoprotein}& 48.95 & 48.95 & 48.95 & 48.95 & 48.95 & 48.95 & 0.00 & 32.86 & 32.86 & 32.86 & 32.86 & 32.86 & 32.86 & 0.00 \\
        \midrule
        FUN & GOA\_Emb & 66.00 & 65.77 & 63.82 & 66.08 & 65.00 & 65.33 & 0.95 & 65.69 & 65.70 & 66.31 & 66.08 & 64.70 & 65.70 & 0.62 \\
        FUN + EXP & GOA\_Emb + Gene2Vec & \textbf{70.22} & \textbf{70.30} & \textbf{70.32} & \textbf{70.73} & \textbf{70.22} & \textbf{70.36} & 0.21 & \textbf{70.13} & \textbf{70.27} & \textbf{70.31} & \textbf{70.67} & \textbf{70.15} & \textbf{70.35} & 0.22 \\
        FUN + SEQ & GoA\_Emb + DNAbert-2 & 68.32 & 68.89 & 69.06 & 68.97 & 69.10 & 68.87 & 0.32 & 68.32 & 68.88 & 69.06 & 68.95 & 69.10 & 69.00 & 0.32 \\
        \bottomrule
    \end{tabular}
    }
\end{table}

\begin{table}[!ht]
    \centering
    \caption{Extra 5-fold results of Single-Cell Type Annotation (Zheng68k\cite{zheng2017massively})}
    \label{appendixtab:extra_results_zheng68k}
    \resizebox{\textwidth}{!}{%
    \begin{tabular}{@{}llccccccc|ccccccc@{}}
        \toprule
        \textbf{Model Type} & \textbf{Model} & \multicolumn{6}{c}{\textbf{Accuracy (\%)}} & \multicolumn{8}{c}{\textbf{F1 Score (\%)}} \\ 
        \cmidrule(lr){3-9} \cmidrule(l){10-16}
        & & \textbf{Fold0} & \textbf{Fold1} & \textbf{Fold2} & \textbf{Fold3} & \textbf{Fold4} & \textbf{Avg} & \textbf{Std} & \textbf{Fold0} & \textbf{Fold1} & \textbf{Fold2} & \textbf{Fold3} & \textbf{Fold4} & \textbf{Avg} & \textbf{Std} \\ 
        \midrule
        EXP & Gene2Vec \cite{du2019gene2vec} & 77.90 & 77.81 & 77.74 & 77.78 & 78.06 & 77.86 & 0.13 & 65.44 & 66.79 & 66.51 & 66.66 & 66.18 & 66.53 & 0.54\\
        SEQ & DNAbert-2 \cite{zhou2023dnabert2} & 74.73 & 74.13 & 74.54 & 74.49 & 73.87 & 74.35 & 0.35 & 60.23 & 59.07 & 60.06 & 59.97 & 58.07 & 59.29 & 0.91\\
        SEQ + FUN & OntoProtein \cite{zhang2022ontoprotein}& 73.51 & 73.38 & 73.69 & 73.59 & 73.73 & 73.58 & 0.14 & 56.88 & 57.80 & 58.75 & 57.95 & 58.94 & 58.36 & 0.83\\
        \midrule
        FUN & GOA\_Emb & 76.84 & 76.80 & 77.41 & 77.18 & 77.12 & 77.07 & 0.25 & 63.80 & 64.37 & 65.50 & 64.77 & 65.32 & 64.99 & 0.69\\
        FUN + EXP & GOA\_Emb + Gene2Vec & \textbf{78.88} & \textbf{78.44} & \textbf{78.84} & \textbf{78.93} & \textbf{78.90} & \textbf{78.80} & 0.20 & \textbf{67.32} & \textbf{66.96} & \textbf{68.02} & \textbf{66.75} & \textbf{67.68} & \textbf{67.35} & 0.52\\
        FUN + SEQ & GoA\_Emb + DNAbert-2 & 77.11 & 76.73 & 77.12 & 77.05 & 77.33 & 77.07 & 0.22 & 63.89 & 63.37 & 65.19 & 64.04 & 64.90 & 64.37 & 0.75\\
        \bottomrule
    \end{tabular}
    }
\end{table}

\begin{table}[htbp]
  \centering
  \caption{Counts and Percentages of Different DB Object Types in GOA}
    \resizebox{\textwidth}{!}{%
    \begin{tabular}{l|lrrrr}
    \toprule
    \textbf{Category} & \textbf{DB Object Type} & \textbf{All GOA} & \textbf{\% (All GOA)} & \textbf{GOA w/o IEA} & \textbf{\% (GOA w/o IEA)} \\
    \midrule
    Protein & protein & 47,384,341 & 59.10\% & 3,664,335 & 49.13\% \\
    \midrule
    Gene & protein\_coding & 11,660,397 & 14.54\% & 2,262,415 & 30.33\% \\
    & protein\_coding\_gene & 619,670 & 0.77\% & 452,056 & 6.06\% \\
    & gene & 174,201 & 0.22\% & 50,313 & 0.67\% \\
    & pseudogene\_with\_CDS & 87 & 0.00\% & 25 & 0.00\% \\
    & pseudogene & 13 & 0.00\% & 3 & 0.00\% \\
    \midrule
    Other & gene\_product & 163,226 & 0.20\% & 61,758 & 0.83\% \\
    \midrule
    RNA & transcript & 20,150,004 & 25.13\% & 939,660 & 12.60\% \\
    & mRNA & 27,764 & 0.03\% & 27,760 & 0.37\% \\
    & miRNA & 1 & 0.00\% & 1 & 0.00\% \\
    & lincRNA & 2 & 0.00\% & 0 & 0.00\% \\
    \midrule
    \textbf{Total} & & 80,179,706 & 100.00\% & 7,458,326 & 100.00\% \\
    \bottomrule
    \end{tabular}%
    }
  \label{appendixtab:DB_object_Type}%
\end{table}%

\begin{table}[ht]
    \centering
    \caption{Distribution of 7,458,326 GOA by Evidence Code (without Inferred from Electronic Annotation (IEA))}
    \label{tab:goa_distribution}
    \resizebox{\textwidth}{!}{%
    \begin{tabular}{@{}l|llrr@{}}
        \toprule
        \textbf{Category} & \textbf{Evidence} & \textbf{Full Name} & \textbf{Number} & \textbf{\% (GOA)} \\
        \textbf{} & \textbf{ Code} & \textbf{} & \textbf{of GOA} & \textbf{} \\
        \midrule
        Author Statement & TAS & Traceable Author Statement & 150,025 & 2.01\% \\
        Evidence Codes & NAS & Non-traceable Author Statement & 49,525 & 0.66\% \\
        \midrule
        Curator Statement & ND & No biological data available & 988,075 & 13.25\% \\
        Evidence Codes & IC & Curator Statement & 46,367 & 0.62\% \\
        \midrule
        Computational Analysis & ISS & Inferred from Sequence or Structural Similarity & 480,486 & 6.44\% \\
        Evidence Codes & ISO & Inferred from Sequence Orthology & 473,456 & 6.35\% \\
        & RCA & Inferred from Reviewed Computational Analysis & 64,672 & 0.87\% \\
        & ISM & Inferred from Sequence Model & 46,514 & 0.62\% \\
        & ISA & Inferred from Sequence Alignment & 16,133 & 0.22\% \\
        & IGC & Inferred from Genomic Context & 3,039 & 0.04\% \\
        \midrule
        Experimental Evidence & IDA & Inferred from Direct Assay & 688,453 & 9.23\% \\
        Evidence Codes & IMP & Inferred from Mutant Phenotype & 592,871 & 7.95\% \\
        & IPI & Inferred from Physical Interaction & 222,050 & 2.98\% \\
        & HDA & Inferred from High Throughput Direct Assay & 200,135 & 2.68\% \\
        & IEP & Inferred from Expression Pattern & 116,862 & 1.57\% \\
        & IGI & Inferred from Genetic Interaction & 76,357 & 1.02\% \\
        & EXP & Inferred from Experiment & 65,651 & 0.88\% \\
        & HMP & Inferred from High Throughput Mutant Phenotype & 54,538 & 0.73\% \\
        & HEP & Inferred from High Throughput Expression Pattern & 1,357 & 0.02\% \\
        & HGI & Inferred from High Throughput Genetic Interaction & 90 & 0.00\% \\
        & HTP & Inferred from High Throughput Experiment & 1 & 0.00\% \\
        \midrule
        Phylogenetically-Inferred  & IBA & Inferred from Biological aspect of Ancestor & 3,120,795 & 41.84\% \\
        Annotations & IKR & Inferred from Key Residues & 874 & 0.01\% \\
        & IBD & Inferred from Biological aspect of Descendant & 0 & 0.00\% \\ 
        & IRD & Inferred from Rapid Divergence & 0 & 0.00\% \\
        \bottomrule
    \end{tabular}
    }
\end{table}

\newgeometry{top=0cm}

\includepdf[pages=-, fitpaper=true]{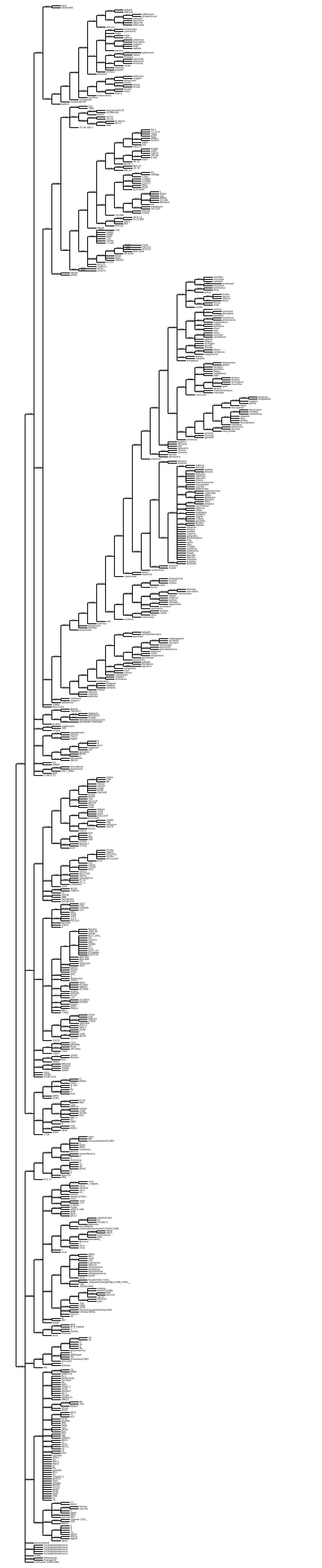}
\restoregeometry

\newpage

\bibliographystyle{plainnat}
\bibliography{ref}